\documentstyle[amssymb,aps,twocolumn,psfig]{revtex}

\begin{document}
\draft
\title{Astrophysical Systems:\\
A model based on the self-similarity scaling postulates.}
\author{L. Velazquez\thanks{%
luisberis@geo.upr.edu.cu}}
\address{Departamento de F\'{i}sica, Universidad de Pinar del R\'{i}o\\
Mart\'{i} 270, esq. 27 de Noviembre, Pinar del R\'{i}o, Cuba. }
\author{F. Guzm\'{a}n\thanks{%
guzman@info.isctn.edu.cu}}
\address{Departamento de F\'{i}sica Nuclear\\
Instituto Superior de Ciencias y Tecnolog\'{i}as Nucleares\\
Quinta de los Molinos. Ave Carlos III y Luaces, Plaza\\
Ciudad de La Habana, Cuba.}
\date{\today}
\maketitle

\begin{abstract}
In the present work, it is developed a formalism to deal with the
macroscopic study of the astrophysical systems, which is based on the
consideration of the exponential self-similarity scaling laws that these
systems exhibit during the realization of the thermodynamic limit. Due to
their scaling laws, these systems are pseudoextensive, since although they
are nonextensive in the usual sense, they can be studied by the
Boltzmann-Gibbs Statistics if an appropriate representation of the integrals
of motion of the macroscopic description is chosen. As example of
application, it is analyzed the system of classical identical particles
interacting via Newtonian interaction. A renormalization procedure is used
in order to perform a well-defined macroscopic description of this system in
quasi-stationary states, since it can not be in a real thermodynamic
equilibrium. Our analysis showed that the astrophysical systems exhibit
self-similarity under the following thermodynamic limit: $E\rightarrow
\infty ,$ $L\rightarrow 0,$ $N\rightarrow \infty ,$ keeping $E/N^{\frac{7}{3}%
}=$const, $LN^{\frac{1}{3}}=$const, where $L$ is the characteristic linear
dimension of the system. It is discussed the effect of these scaling laws in
the dynamical properties of the system. In a general way, our solution
exhibits the same features of the Antonov problem: the existence of the
gravitational collapse at low energies as well as a region with a negative
heat capacity.
\end{abstract}

\pacs{PACS numbers: 05.20.-y; 05.70.-a}

\section{Introduction}

Traditional thermodynamics is not able to describe the astrophysical
systems. They do not fulfill the additivity and homogeneity conditions,
indispensable requirements for the good performance of this formulation.

In the last years it has been devoted so much effort to the extension of the
Thermodynamics to the study of the nonextensive systems. In this context,
the astrophysical systems have received a special attention.

Among the remarkable new results obtained in the present frame, it can be
mentioned the application of the so popular Tsallis' nonextensive statistics 
\cite{tsal} to the analysis of astrophysical systems. In this approach it
has been {\em justified} the application of the polytropic models \cite
{pla1,pla2}, which have been extensively used in the descriptions of such a
systems \cite{chand,kip}:

\begin{equation}
p=C\rho ^{\gamma }\text{ , }\gamma >1\text{,}
\end{equation}
where $p$ is the pressure, $\rho $, the particles density, and $\gamma $,
the polytropic index.

However, in our opinion, in those works it has been put off the microscopic
justification of the polytropic models to the applicability of the Tsallis'
theory. The nonextensive statistics is not a completely satisfactory
formulation, due to the all theory dependence on the entropic index, $q$,
which is a parameter representing a measure of the degree of nonextensivity
of the system. In spite of the attractiveness of this formulation, the
theory is not able to determine univocally the value of the entropic index,
at least, in the context of the equilibrium Thermodynamics, so that, it must
be appealed to the experiment or computational simulations in order to
precise it. Some evidences aim that the entropy index could be obtained
throughout the sensitivity of the system to the initial conditions and the
relaxation properties towards equilibrium \cite{dyn,dyn2,dyn3,dyn4,dyn5,reis}%
.

Alternative approaches have been proposed using the microcanonical ensemble.
Although it can not be assured its application to any Hamiltonian system%
\footnote{%
Apparentely the microcanonical ensemble is {\em only well-defined}
statistical ensemble, whose justification ordinarily is based on the chaotic
properties of the trayectories for a generic non-integrable system when it
is overcome a few tens of degrees of freedom. However, there are some
computational evidences in dynamical studies of some nonextensive systems in
which the mixing time is extremely long, and diverges with the increasing of
system degrees of freedom. In some of these cases is also observed the
existence of anomalous {\em quasi-stationary states} which can not be
described using the microcanonical ensemble. Only after a long trasient
time, the system abandons these singular states and starts slowly
approaching to the equilibrium dictated by the microcanonical ensemble. In
this case it is shown that the infinite time limit $\left( t\rightarrow
\infty \right) $ necessary to the stablishment of the equilibrium described
with microcanonical ensemble does not commute with the thermodynamic limit $%
\left( N\rightarrow \infty \right) $, since the second is performed before
the first, the quasi-stationary state becomes the real equilibrium state of
the system. The interested reader may check the following reference for
details: Phys. Rev. E 64, 056134 (2001).}, the Thermo-statistics could be
justified starting from this ensemble without invoking anything outside the
Mechanics. This description is applicable to many situations in which the
canonical description fails, allowing us to determine the necessary
conditions for the applicability of any generalized canonical ensemble.

Starting from this viewpoint, in the ref.\cite{vel1} it was addressed a
generalization of the extensive postulates of the traditional Thermodynamics
in order to extend its application to the nonextensive systems. According to
our proposition, this objective could be carried out taking into
consideration the {\em self-similarity scaling postulates}: the equivalence
of the microcanonical ensemble with a generalized canonical one during the
realization of the thermodynamic limit throughout of the {\bf %
self-similarity scaling properties} of the system fundamental physical
observables: the behavior of integrals of motion, the external parameters
and the accessible volume of the microcanonical description with the
increasing of the system degrees of freedom.

So far, these postulates have been applied to the analysis of the necessary
conditions for the validity of two statistical formulations: in the ref.\cite
{vel2}, to the {\em microcanonical thermostatistics }of D. H. E. Gross \cite
{gro1,gro2}, as well as in the ref.\cite{vel3}, to the Tsallis' nonextensive
statistics \cite{tsal}.

The first is a theory based on the consideration of the microcanonical
ensemble with the assumption of the Boltzmann's definition of entropy:

\begin{equation}
S_{B}=\ln W,  \label{be}
\end{equation}
his famous gravestone epithaph in Vienna. Since the Boltzmann's entropy does
not demand the realization of the thermodynamic limit this formulation is
applicable to some small and mesoscopic systems. The thermodynamic formalism
of this theory has been defined in order to be equivalent to the traditional
one when it is applied to extensive systems. That is the reason why this
theory is appropriate to the macroscopic description of those systems
becoming extensive when the thermodynamic limit is invoked although they are
not found in the thermodynamic limit.

In the ref.\cite{vel2} it was shown that the Gross's theory is also
applicable to all those Hamiltonian systems exhibiting {\em exponential
self-similarity scaling laws} in the thermodynamic limit. We called these
system as {\em pseudoextensive }when{\em \ }$N\rightarrow \infty $, since in
spite of they are nonextensive in the usual sense, they can be tried by
means of the Boltzmann-Gibbs' statistics if {\em an appropriate
representation of the integrals of motion is chosen}. Many of the systems
found in the real world belong to the class of the pseudoextensive systems,
since it is enough an additive kinetic part in the Hamiltonian of the system
for exhibiting this kind of scaling laws in the thermodynamic limit.

In the ref.\cite{vel3} was shown that the Tsallis' statistics is appropriate
to the macroscopic description of those Hamiltonian nonextensive systems
exhibiting a {\em potential self-similarity scaling laws} in the
thermodynamic limit. In our approach many details of this formalism
naturally appear, starting only from the Mechanics under the consideration
of this kind of scaling laws, i.e., the {\em q}-expectation values, the {\em %
q}-generalization of the Legendre's Transformation (see for example in refs. 
\cite{Abe1,Fran}).

The above reasons allow us to consider that the astrophysical systems belong
to the class of the pseudoextensive systems, and therefore, it is also
justified the application of the Boltzmann-Gibbs' statistics to the study of
these systems, at least, when the phase transitions are not present.

The Tsallis' statistics is expected to describe systems exhibiting potential
distributions, that is, systems with fractal characteristics. In the ref. 
\cite{han} it is suggested that the Tsallis' potential distribution fitted
very well the differential energy distribution of the dark matter in halos
obtained by a numerical simulation.\ Similar analysis have been carried out
in ref.\cite{fa} by Fa \& Pedron for the elliptical galaxies. However, the
fractals properties can be also found in the context of Boltzmann-Gibbs'
statistics, for example, in the Michie-King models for globular clusters 
\cite{bin} (see ref.\cite{inag} for review), or the models proposed by
Stiavelli \& Bertin \cite{st}, Hjorth \& Madsen \cite{hjor} for elliptical
galaxies, and others \cite{ing}.\ These models lead to composite
configurations with an isothermal core and a polytropic envelope and can not
be justified by the Tsallis' generalized Thermodynamics. Many authors state
that these models are more appropriate for the description of those
astrophysical objects (see for example in ref.\cite{chava1}).

In the present work it is pretended to reconsider the thermo-statistical
description of the astrophysical system, but this time, taking into
consideration the self-similarity scaling postulates \cite{vel1} in order to
find the necessary conditions for the validity of the generalized canonical
ensemble (Boltzmann-Gibbs) in the thermodynamic limit.

\section{Microcanonical description.}

\subsection{ \ \ Microcanonical Mean Field Approximation.}

Let ${\cal S}$ be a Hamiltonian system composed by a huge number $N$ of
identical particles, which interact among them by means of short-range and
long-range forces simultaneously. Let us also consider that the
characteristic linear dimension of the system is comparable with the
effective radio of the long-range interactions, but it is extremely large in
comparison with the effective radio of the short-range interactions.

The above conditions allow us to speak about two {\em small} scales in the
system:

\begin{itemize}
\item  {\em Microscopic scale}: linear dimensions comparable with the
effective radio of the short-range interactions.

\item  {\em Local scale}: very large linear dimensions in regard to the
microscopic scale, but extremely short in comparison with the effective
radio of the long-range interactions.
\end{itemize}

Let us consider a partition of accesible physical space in not overlapped
cells, $\left\{ c_{k}\right\} $, whose {\em characteristic linear dimensions
correspond to a local scale}. It is easy to show that the {\em N}-body phase
space integration can be descomposed in the following way:

\begin{equation}
\frac{1}{N!}%
\mathrel{\mathop{\int }\limits_{{\cal X}_{N}}}%
dX_{N}\equiv 
\mathrel{\mathop{\prod }\limits_{k}}%
\mathrel{\mathop{\stackrel{N}{\sum }}\limits_{n_{k}=0}}%
\widehat{{\cal O}}\left[ {\cal X}_{n_{k}}^{\left( k\right) }\right] \delta
^{\left( e\right) }\left( N-%
\mathrel{\mathop{\sum }\limits_{k}}%
n_{k}\right) ,  \label{partition}
\end{equation}
where it have been taken into account all the possible configurations for
distributing $N$ identical particles in the cells. In this expression, $%
{\cal X}_{N}$ represents the {\em N}-body phase space, while ${\cal X}%
_{n_{k}}^{\left( k\right) }$ represents {\em n}$_{k}$-body phase space whose
accesible physical space have been limited to the {\em k-th} cell, $c_{k}$.
In addition, it has been introduced the following integral operator $%
\widehat{{\cal O}}\left[ {\cal X}_{n}^{\left( k\right) }\right] $:

\begin{equation}
\widehat{{\cal O}}\left[ {\cal X}_{n}^{\left( k\right) }\right] =\left\{ 
\begin{array}{c}
\frac{1}{n!}%
\mathrel{\mathop{\int }\limits_{{\cal X}_{n}^{\left( k\right) }}}%
dX_{n},\text{ if }n\neq 0, \\ 
1,\text{ if }n=0,
\end{array}
\right.
\end{equation}
as well as the function $\delta ^{\left( e\right) }\left( n\right) \equiv
\delta _{0n}$, which assures the particles number conservation. A derivation
of the Eq.(\ref{partition}) appears in the appendix A.

The enormous scale separation among the effects of the short and long-range
interactions supports the validity of the {\em spacial adiabatic
approximation}. Let us consider those configurations in which each cell $%
c_{k}$ contains a very large number of particles $n_{k}$ and let us denote
this subsystem of particles by ${\cal S}_{k}$. The physical quantities
characterizing the long-range interactions, that is, the long-range
interacting fields $\phi \left( {\bf r}\right) $, almost do not vary at the
spacial region occupied by the cell due to its linear dimension. It means
that the long-range interactions almost do not distinguish the internal
structure of the subsystem enclosed by the cell, so that, these interactions
are only effective for the subsystem as a whole.

Therefore, the contribution of long-range interactions to the system total
energy can be approximated by functional terms of the mean values of the
long-range interacting fields at the region of each cell, $\phi _{k}$, as
well as of certain collective quantities $q_{k}$ characterizing the
subsystems $\left\{ {\cal S}_{k}\right\} $, like the mean values of {\em %
particles density}, {\em magnetization density, etc.} Similarly, the mean
values of the long-range interacting fields can be determined from the mean
values of some collective quantities of the subsystems $\left\{ {\cal S}%
_{k}\right\} $. On the other hand, due also to the linear dimensions of the
cells, it can be neglated the short-range interactions among the particles
beloging to different cells.

Taking into account all the above exposed, the system Hamiltonian can be
approximately expressed as follows:

\begin{equation}
H\simeq 
\mathrel{\mathop{\sum }\limits_{k}}%
h_{int}^{\left( n_{k}\right) }\left( X_{n_{k}};\phi _{k}\right) +{\cal V}%
_{loc}\left( \phi _{k},q_{k}\right) \text{,}  \label{energy integral}
\end{equation}
where $h_{int}^{\left( n_{k}\right) }\left( X_{k};\phi _{k}\right) $ is the
internal energy of the $n_{k}$-body subsystem ${\cal S}_{k}$ enclosed in the
cell $c_{k}$, which only involves the kinetic energy of the particles as
well as the contribution of the short-range interactions among them, being $%
X_{n_{k}}$ their microscopic degrees of freedom. Here, it is also included a
parametric dependence of the internal energy of the fields $\phi _{k}$ in
order to take into account a possible influence of these fields on the
internal configurations of the subsystem\ ${\cal S}_{k}$. On the other hand, 
${\cal V}_{loc}$ is a local term of energy containing the contribution of
the subsystem ${\cal S}_{k}$ as a whole, which only involves its effective
interactions with the long-range interacting fields $\phi _{k}$. As already
mentioned, the fields $\phi _{k}$ are determined through a determined
functionals of certain set local quantities, $q\equiv \left\{ q_{k}\right\} $%
, which characterize the local subsystems:

\begin{equation}
\phi _{k}={\cal F}_{k}\left( q\right) \text{.}
\end{equation}

As it could be seen, the effective long-range interacting fields $\phi _{k}$
at the {\em k-th} cell can be considered as external parameters for the
subsystem ${\cal S}_{k}$. Therefore, each subsystem ${\cal S}_{k}$ can be
considered as {\em locally extensive}. Due to the presence of the long-range
interactions, the system ${\cal S}=$ $%
\mathrel{\mathop{\bigcup }\limits_{k}}%
{\cal S}_{k}$ is nonextensive: the quantities characterizing the subsystems
as a whole vary during the continuous passage among the cells.

There are systems in which the energy is not the only one integral of motion
determining their macroscopic description (for example: the astrophysical
systems), so that, it could be considered other integrals of motion, like
total angular momentum \cite{bin}. In such as cases, it is well-known that
the total angular momentum admits a descomposition similar to the expression
of the energy in the Eq.(\ref{energy integral}):

\begin{equation}
{\bf M}=%
\mathrel{\mathop{\sum }\limits_{k}}%
{\bf m}_{int}^{\left( n_{k}\right) }\left( \widetilde{X}_{n_{k}}\right) +\mu
_{k}{\bf r}_{k}\times {\bf v}_{k},
\end{equation}
where ${\bf r}_{k}$ and ${\bf v}_{k}$ are the position and velocity of the
mass center of the subsystem ${\cal S}_{k}$, $\mu _{k}$ is its total mass,
while ${\bf m}_{int}^{\left( n_{k}\right) }\left( \widetilde{X}%
_{n_{k}}\right) $ and $\widetilde{X}_{n_{k}}$ are the internal angular
momentum and the microscopic degrees of freedom of the subsystem in its own
mass center frame. The consideration of the collective motion of the
subsystem ${\cal S}_{k}$ leads to add a kinetic term $\frac{1}{2}\mu _{k}%
{\bf v}_{k}^{2}$ to the sum in the Eq.(\ref{energy integral}) as well as to
substitute $X_{n_{k}}$ by $\widetilde{X}_{n_{k}}$.

Thus, the integrals of motion involved in the macroscopic description of the
system can be represented in the following form after the considerations
assumed above:

\begin{equation}
I\simeq 
\mathrel{\mathop{\sum }\limits_{k}}%
{\cal I}_{internal}^{\left( n_{k}\right) }\left( X_{n_{k}};\phi _{k}\right) +%
{\cal J}_{collective}\left( \phi _{k},q\right) ,
\end{equation}
where ${\cal I}_{internal}^{\left( n_{k}\right) }\left( X_{n_{k}};\phi
_{k}\right) $ represents the internal contribution from the microscopic
degrees of freedom of the subsystem ${\cal S}_{k}$, while ${\cal J}%
_{collective}\left( \phi _{k},q_{k}\right) $ is the contribution of its {\em %
collective} degrees of freedom at local level.

According with what was exposed above, the accessible volume of the system
in the microcanonical ensemble is approximately given by:

\[
W\left( I,N\right) \simeq 
\mathrel{\mathop{\prod }\limits_{k}}%
\int d\phi _{k}\left( 
\mathrel{\mathop{\stackrel{N}{\text{ }\sum }}\limits_{n_{k}=0}}%
\widehat{{\cal O}}\left[ {\cal X}_{n_{k}}^{\left( k\right) }\right] \delta %
\left[ \phi _{k}-{\cal F}_{k}\left( q\right) \right] \times \right. 
\]
\[
\times \left. \delta ^{\left( e\right) }\left( N-%
\mathrel{\mathop{\sum }\limits_{k}}%
n_{k}\right) \right) \times 
\]

\begin{equation}
\times \delta \left[ I-%
\mathrel{\mathop{\sum }\limits_{k}}%
{\cal I}_{internal}^{\left( n_{k}\right) }\left( \widetilde{X}_{n_{k}};\phi
_{k}\right) +{\cal J}_{collective}\left( \phi _{k},q_{k},{\bf v}_{k}\right) %
\right] ,  \label{de1}
\end{equation}
where it has been explicitly introduced the dependence of the collective
term of the mass center velocity, in order to take into account the
collective motion of the subsystems. Using the identity:

\begin{equation}
\int di_{k}d{\bf v}_{k}\delta \left[ i_{k}-{\cal I}_{internal}^{\left(
n_{k}\right) }\left( \widetilde{X}_{n_{k}};\phi _{k}\right) \right] \delta %
\left[ {\bf v}_{k}-\frac{1}{\mu _{k}}{\bf P}_{k}\right] \equiv 1,
\end{equation}
where ${\bf P}_{k}$ is the total linear momentum of the subsystem ${\cal S}%
_{k}$, the Eq.(\ref{de1}) can be rewritten introducing the Boltzmann's
entropy for each local subsystems ${\cal S}_{k}$, $S_{B}(i_{k},n_{k};\phi
_{k})$:

$\exp \left[ S_{B}(i_{k},n_{k};\phi _{k})\right] =$%
\begin{equation}
\frac{1}{n_{k}!}%
\mathrel{\mathop{\int }\limits_{{\cal X}_{n_{k}}^{\left( k\right) }}}%
dX_{n_{k}}\delta \left[ i_{k}-{\cal I}_{internal}^{\left( n_{k}\right)
}\left( X_{n_{k}};\phi _{k}\right) \right] \delta \left[ {\bf P}_{k}\right] 
\text{,}
\end{equation}
where $i_{k}$ is the internal contribution of the subsystem to the total
integrals of motion of the system ${\cal S}$. Thus, the accessible volume of
the system can be computed from 
\[
W\left( I,N\right) \simeq 
\mathrel{\mathop{\prod }\limits_{k}}%
\mathrel{\mathop{\stackrel{N}{\text{ }\sum }}\limits_{n_{k}=0}}%
\int di_{k}d\phi _{k}d{\bf v}_{k}\delta \left[ \phi _{k}-{\cal F}_{k}\left(
q\right) \right] \times 
\]
\[
\times \exp \left[ 
\mathrel{\mathop{\sum }\limits_{k}}%
S_{B}(i_{k},n_{k};\phi _{k})\right] \delta ^{\left( e\right) }\left( N-%
\mathrel{\mathop{\sum }\limits_{k}}%
n_{k}\right) \times 
\]
\begin{equation}
\times \delta \left[ I-%
\mathrel{\mathop{\sum }\limits_{k}}%
i_{k}+{\cal J}_{collective}\left( \phi _{k},q_{k},{\bf v}_{k}\right) \right]
.  \label{de2}
\end{equation}
where the physical quantities $q_{k}$ are determined from the $i_{k}\ $by
means of certain functional dependencies:

\begin{equation}
q_{k}=f_{q}\left( i_{k},n_{k},{\bf v}_{k}\right) \text{.}
\end{equation}

Developing the {\em continuum limit}:

\begin{equation}
N\rightarrow \infty ,\text{ and\ \ }\mu \left( c_{j}\right) /\left( 
\mathrel{\mathop{\sum }\limits_{k}}%
\mu \left( c_{k}\right) \right) \rightarrow 0,
\end{equation}
where $\mu \left( c_{k}\right) $ is the physical volume of the {\em k-th}
cell, the subsystem ${\cal S}_{k}$ is locally perceived as a fluid. Thus, it
is obtained finally {\em the microcanonical mean field approximation} (MFA)%
{\em :}

$W\left( I,N\right) \simeq W_{MFA}\left( I,N\right) =$

\[
C\int {\cal D}\varrho \left( {\bf r}\right) {\cal D}\rho \left( {\bf r}%
\right) {\cal D}\phi \left( {\bf r}\right) {\cal D}{\bf v}\left( {\bf r}%
\right) \delta \left\{ \phi \left( {\bf r}\right) -{\cal F}_{\phi }\left[ 
{\bf r};\varrho ,\rho ,{\bf v}\right] \right\} \times 
\]

\[
\times \exp \left[ \int d^{3}{\bf r}\text{ }s\left[ \varrho \left( {\bf r}%
\right) ,\rho \left( {\bf r}\right) ;\phi \left( {\bf r}\right) \right] %
\right] \delta \left( N-\int d^{3}{\bf r}\text{ }\rho \left( {\bf r}\right)
\right) \times 
\]

\begin{equation}
\times \delta \left[ I-\int d^{3}{\bf r}\text{ }\varrho \left( {\bf r}%
\right) +\vartheta \left[ {\bf r;}\phi \left( {\bf r}\right) ,\varrho \left( 
{\bf r}\right) ,\rho \left( {\bf r}\right) ,{\bf v}\left( {\bf r}\right) %
\right] \right] ,
\end{equation}
where $\rho \left( {\bf r}\right) $, $\varrho \left( {\bf r}\right) $, $%
\vartheta \left[ {\bf r,}\phi \left( {\bf r}\right) ,\varrho \left( {\bf r}%
\right) ,\rho \left( {\bf r}\right) ,{\bf v}\left( {\bf r}\right) \right] $,
and $s\left[ \varrho \left( {\bf r}\right) ,\rho \left( {\bf r}\right) ;\phi
\left( {\bf r}\right) \right] $ are respectively: the particles density, the
densities of the internal and collective contributions to the total
integrals of motions of the system, and the entropy density of the fluid at
the neighborhood of the point ${\bf r}$. $C$ is an unimportant constant
which appears as consequence of the continuous limit and can be ignored.

\subsection{The case of the astrophysical systems.}

Let us apply the microcanonical mean field approximation to the analysis of
an astrophysical system composed by a static fluid of identical particles,
whose macroscopic state is only determined by its total mass\ (number of
particles) and the energy. Let $\varepsilon \left( {\bf r}\right) $, $\rho
\left( {\bf r}\right) $ and $s\left[ \varepsilon ,\rho ;\phi \right] $ be
respectively the internal energy, particle, and entropy densities of the
system neighborhood of the point ${\bf r}$, where $\phi $ is the
self-gravitating Newtonian potential. In this case the microcanonical mean
field approximation is written as follows:

\[
W_{MFA}\left( E,N\right) =\int {\cal D}\varepsilon \left( {\bf r}\right) 
{\cal D}\rho \left( {\bf r}\right) {\cal D}\phi \left( {\bf r}\right) \delta
\left\{ \phi \left( {\bf r}\right) -{\cal L}\left[ {\bf r};\rho \right]
\right\} 
\]

\[
\times \exp \left[ \int d^{3}{\bf r}\text{ }s\left[ \varepsilon \left( {\bf r%
}\right) ,\rho \left( {\bf r}\right) ;\phi \left( {\bf r}\right) \right] %
\right] \delta \left[ N-\int d^{3}{\bf r}\text{ }\rho \left( {\bf r}\right) %
\right] \times 
\]

\begin{equation}
\times \delta \left[ E-\int d^{3}{\bf r}\text{ }{\cal H}\left[ \varepsilon
\left( {\bf r}\right) ,\rho \left( {\bf r}\right) ,\phi \left( {\bf r}%
\right) \right] \right] \text{,}  \label{aq}
\end{equation}
where ${\cal H}\left[ \varepsilon \left( {\bf r}\right) ,\rho \left( {\bf r}%
\right) ,\phi \left( {\bf r}\right) \right] $ is the Hamiltonian density of
the system, which is given by:

\begin{equation}
{\cal H}\left[ \varepsilon ,\rho ,\phi \right] =\frac{1}{8\pi G}\left( {\bf %
\nabla }\phi \right) ^{2}+m\rho \phi +\varepsilon \text{.}  \label{halm}
\end{equation}
In the above definition: $G$ is the Newton's constant and $m$ is the
particles mass. Finally, ${\cal L}\left[ {\bf r};\rho \right] $ is the
functional:

\begin{equation}
{\cal L}\left[ {\bf r};\rho \right] =m\int d^{3}{\bf r}^{\prime }\text{ }%
g\left( {\bf r,r}^{\prime }\right) \rho \left( {\bf r}^{\prime }\right) 
\text{,}  \label{gs}
\end{equation}
which determines the spacial configuration of the Newtonian potential $\phi
\left( {\bf r}\right) $ at a given spacial configuration of the particles
density, $\rho \left( {\bf r}\right) $, where $g\left( {\bf r,r}^{\prime
}\right) $ is the Green's function of the Poisson's problem:

\begin{equation}
\Delta G\left( {\bf r,r}^{\prime }\right) =4\pi G\delta \left( {\bf r-r}%
^{\prime }\right) \text{.}
\end{equation}
In the tridimensional case:

\begin{equation}
g\left( {\bf r,r}^{\prime }\right) =-\frac{G}{\left| {\bf r-r}^{\prime
}\right| }\text{.}  \label{tgf}
\end{equation}

Using the Fourier's representation of the Dirac's delta function:

\begin{equation}
\delta \left( x-x^{\prime }\right) =\int \frac{dk}{2\pi }\exp \left[ z\left(
x-x^{\prime }\right) \right] \text{,}
\end{equation}
where $z=\beta +ik$, with $\beta \in {\bf R}$, the Eq.(\ref{aq}) can be
rewritten as follows:

\begin{equation}
W_{MFA}\left( E,N\right) =\stackrel{+\infty }{%
\mathrel{\mathop{\int }\limits_{-\infty }}%
}\frac{dk}{\left( 2\pi \right) ^{2}}{\cal Z}\left[ z_{1},N\right] \exp \left[
z_{1}E\right] .
\end{equation}
The functional ${\cal Z}\left[ z_{1},N\right] $, with argument $z_{1}=\beta
+ik$, is given by:

\begin{equation}
{\cal Z}\left[ z_{1},N\right] =\stackrel{+\infty }{%
\mathrel{\mathop{\int }\limits_{-\infty }}%
}\frac{ds}{2\pi }\exp \left[ z_{2}N\right] \aleph \left( z_{1},z_{2}\right) ,
\end{equation}
with $z_{2}=\mu +is$ , $\mu \in {\bf R}$, and the function $\aleph \left(
z_{1},z_{2}\right) $ is expressed as follows:

\[
\aleph \left( z_{1},z_{2}\right) =\int {\cal D}\varepsilon \left( {\bf r}%
\right) {\cal D}\rho \left( {\bf r}\right) {\cal D}\phi \left( {\bf r}%
\right) {\cal D}J\left( {\bf r}\right) \times 
\]
\begin{equation}
\times \exp \left[ -H\left( \varepsilon ,\rho ,\phi ,\omega
;z_{1},z_{2}\right) \right] \text{.}
\end{equation}
The functional $H\left( \varepsilon ,\rho ,\phi ,\omega ;z_{1},z_{2}\right) $
is expressed by:

\[
H\left( \varepsilon ,\rho ,\phi ,\omega ;z_{1},z_{2}\right) =\int d^{3}{\bf r%
}\ z_{1}{\cal H}\left[ \varepsilon \left( {\bf r}\right) ,\rho \left( {\bf r}%
\right) ,\phi \left( {\bf r}\right) \right] + 
\]

\begin{equation}
+z_{2}\rho \left( {\bf r}\right) -\omega \left( {\bf r}\right) \left\{ \phi
\left( {\bf r}\right) -{\cal L}\left[ {\bf r},\rho \right] \right\} -s\left[
\varepsilon \left( {\bf r}\right) ,\rho \left( {\bf r}\right) ;\phi \left( 
{\bf r}\right) \right] \text{,}  \label{HF}
\end{equation}
where $\omega \left( {\bf r}\right) =j\left( {\bf r}\right) +iJ\left( {\bf r}%
\right) $. The auxiliary field $J\left( {\bf r}\right) $ allows us the
Fourier's representation of the delta functional of the Newtonian potential, 
$\phi $, and $j\left( {\bf r}\right) $ is an arbitrary real function.

As it was previously pointed out, the astrophysical systems can be
considered as {\em pseudoextensive} \cite{vel2}: they exhibit an exponential
self-similarity scaling laws in the thermodynamic limit \cite{vel1,vel2},
the limit of many particles. As consequence of this behavior, they can be
dealt with the usual Boltzmann-Gibbs' Statistics, only if an appropriate
representation of the integrals of motion determining their macroscopic
state is chosen and if they do not present first-order phase transitions. In
order to select a correct representation for the integrals of motion, the
self-similarity scaling laws of the system must be found.

It is demanded the following symmetry:

\begin{equation}
\left. 
\begin{tabular}{ll}
$N\rightarrow \alpha N,$ & $\mu \rightarrow \mu ,$ \\ 
$\beta E\rightarrow \alpha \beta E,$ & ${\cal P}\rightarrow \alpha {\cal P}$%
\end{tabular}
\right\} \Rightarrow S_{B}\rightarrow \alpha S_{B}\text{,}
\end{equation}
in order to make equivalent the microcanonical with the canonical
description when $N\rightarrow \infty $, where ${\cal P}$ is the Planck's
potential:

\begin{equation}
{\cal P}\left( \beta ,N\right) =-\ln {\cal Z}\left[ \beta ,N\right] \text{,}
\end{equation}
and $S_{B}$, the Boltzmann's entropy, Eq.(\ref{be}). The above requests
characterize the system as a whole. However, the description of the system
is performed at a local level, and therefore, the scaling laws must be
determined for the local fields: $\varepsilon \left( {\bf r}\right) $, $\rho
\left( {\bf r}\right) $, $\phi \left( {\bf r}\right) $, $j\left( {\bf r}%
\right) $, $s\left[ \varepsilon ,\rho ;\phi \right] $, the parameter $\beta $%
, as well as the scaling law of the spacial coordinate ${\bf r}$. Let us
consider the following scaling laws:

\begin{equation}
\begin{tabular}{llll}
${\bf r}\rightarrow \alpha ^{c}{\bf r,}$ & $\beta \rightarrow \alpha ^{\pi
}\beta ,$ & $\varepsilon \rightarrow \alpha ^{\chi }\varepsilon ,$ & $\phi
\rightarrow \alpha ^{\eta }\phi ,$ \\ 
$\mu \rightarrow \mu ,$ & $j\left( {\bf r}\right) \rightarrow j\left( {\bf r}%
\right) ,$ & $\rho \rightarrow \alpha ^{\varkappa }\rho ,$ & $%
s_{0}\rightarrow \alpha ^{m}s_{0},$%
\end{tabular}
\label{els}
\end{equation}
where $c$, $\pi $, $\chi $, $\eta $, $\varkappa $ and $m$, are certain real
scaling exponent constants. From the analysis of the Eq.(\ref{HF}) it is
demanded the following relations in order to satisfy the {\em homogeneous
scaling} of the functional $H\left( \varepsilon ,\rho ,\phi ,\omega
;z_{1},z_{2}\right) $:

\[
\begin{tabular}{lll}
$3c+m=1,$ & $3c+\varkappa =1,$ & $3c+\pi +\chi =1,$%
\end{tabular}
\]

\begin{equation}
\begin{tabular}{ll}
$\eta -2c=\varkappa ,$ & $2\left( \eta -c\right) =\chi $.
\end{tabular}
\end{equation}
The solution of the above equation system is given by:

\[
\begin{tabular}{lll}
$m=\varkappa ,$ & $\eta =\frac{\varkappa +2}{3},$ & $\pi =-\eta ,$%
\end{tabular}
\]

\begin{equation}
\begin{tabular}{ll}
$c=\frac{1-\varkappa }{3},$ & $\chi =\frac{4\varkappa +2}{3}.$%
\end{tabular}
\label{elp}
\end{equation}

The scaling law for the total energy is given by:

\begin{equation}
E\rightarrow \alpha ^{\tau }E\text{,}  \label{esr}
\end{equation}
where:

\begin{equation}
\tau =\frac{\varkappa +5}{3}\text{,}
\end{equation}
and therefore, an appropriate selection of the representation of the
integrals of motion is:

\begin{equation}
I=\left( {\cal E},N\right) \text{ with }{\cal E}\equiv E/N^{\eta }\text{.}
\end{equation}
In this case, the correct Legendre's transformation between the
thermodynamic potential is given by:

\begin{equation}
S_{B}\left( {\cal E},N\right) \simeq \beta _{o}{\cal E}-{\cal P}\left( \beta
_{o},N\right) ,  \label{legen}
\end{equation}
with:

\begin{equation}
\beta _{o}=\beta N^{\eta }\text{ and }{\cal E}=\frac{\partial }{\partial
\beta _{o}}{\cal P}\left( \beta _{o},N\right) .  \label{cer}
\end{equation}

The above scaling laws depend on the parameter $\varkappa $, so that, they
are only specified when the microscopic model for the fluid is assumed. When
the thermodynamic limit is performed in the generalized canonical ensemble
it will found that the system will carry out more probably those
configurations minimizing the functional $H\left( \varepsilon ,\rho ,\phi
,\omega ;z_{1},z_{2}\right) $. Thus, it is arrived to an {\em equilibrium} 
{\em mean field theory}. These configurations are obtained by solving the
following equations:

\[
\frac{\delta H\left( \varepsilon ,\rho ,\phi ,j;\beta ,\mu \right) }{\delta
\varepsilon \left( {\bf r}\right) }=0\text{, \ \ \ \ \ \ \ }\frac{\delta
H\left( \varepsilon ,\rho ,\phi ,j;\beta ,\mu \right) }{\delta \rho \left( 
{\bf r}\right) }=0\text{,} 
\]

\begin{equation}
\text{ and \ }\frac{\delta H\left( \varepsilon ,\rho ,\phi ,j;\beta ,\mu
\right) }{\delta \phi \left( {\bf r}\right) }=0\text{, }  \label{cmin}
\end{equation}
imposing the contrains:

\begin{equation}
\frac{\delta H\left( \varepsilon ,\rho ,\phi ,j;\beta ,\mu \right) }{\delta
j\left( {\bf r}\right) }=0\text{, }\frac{\delta H\left( \varepsilon ,\rho
,\phi ,j;\beta ,\mu \right) }{\delta \mu }=N\text{,}  \label{ccd}
\end{equation}
which are related with the conservation of the particles number and the
consideration of the Poisson's equation for the Newtonian potential $\phi $.
The maximization takes place when it is guarantied the positive definition
of the functional matrix:

\begin{equation}
D_{ij}\left( {\bf r}^{\prime },{\bf r}\right) =\left. \frac{\delta ^{2}H}{%
\delta f_{i}\left( {\bf r}^{\prime }\right) \delta f_{j}\left( {\bf r}%
\right) }\right| _{f=f_{s}}\text{ },  \label{mf}
\end{equation}
with $f_{i=1,2}=\left( \varepsilon ,\rho \right) $, where the subindex $s$
represents the solution for the minimization conditions, Eqs.(\ref{cmin})
and (\ref{ccd}). From the conditions given in the Eq.(\ref{cmin}) are
derived the following relations:

\begin{eqnarray}
\beta &=&\frac{\partial }{\partial \varepsilon _{s}}s\left( \varepsilon
_{s},\rho _{s};\phi _{s}\right) , \\
\text{ }\mu &=&\frac{\partial }{\partial \rho _{s}}s\left( \varepsilon
_{s},\rho _{s};\phi _{s}\right) -\beta m\phi _{s}-{\cal L}\left( {\bf r}%
,j_{s}\right) \text{,}
\end{eqnarray}
the conditions for thermodynamic equilibrium along the volume of the system,
as well as the structure equation:

\begin{equation}
\Delta \phi _{s}=4\pi G\left[ m\rho _{s}-\beta ^{-1}\left( j_{s}+\frac{%
\partial }{\partial \phi _{s}}s\left( \varepsilon _{s},\rho _{s};\phi
_{s}\right) \right) \right] \text{.}
\end{equation}
From the validity of the Poisson's equation:

\begin{equation}
\Delta \phi _{s}=4\pi Gm\rho _{s}\text{,}
\end{equation}
the functional dependency for the auxiliary field $j_{s}\left( {\bf r}%
\right) $ is derived:

\begin{equation}
j_{s}=-\frac{\partial }{\partial \phi _{s}}s\left( \varepsilon _{s},\rho
_{s};\phi _{s}\right) \text{.}
\end{equation}

It could be also used another alternative choice for the Hamiltonian of the
system given in the Eq.(\ref{HF}), :

\begin{equation}
{\cal H}\left( \varepsilon ,\rho ,\phi \right) =\varepsilon +\frac{1}{2}%
m\rho \phi \text{.}  \label{ah}
\end{equation}
In this case equilibrium conditions lead to the following relations:

\begin{eqnarray}
\beta &=&\frac{\partial }{\partial \varepsilon _{s}}s\left( \varepsilon
_{s},\rho _{s};\phi _{s}\right) , \\
\mu &=&\frac{\partial }{\partial \rho _{s}}s\left( \varepsilon _{s},\rho
_{s};\phi _{s}\right) -\frac{1}{2}\beta m\phi _{s}-{\cal L}\left( {\bf r}%
,j_{s}\right) \text{,} \\
j_{s} &=&\frac{1}{2}\beta m\phi _{s}-\frac{\partial }{\partial \phi _{s}}%
s\left( \varepsilon _{s},\rho _{s};\phi _{s}\right) .
\end{eqnarray}

Taking into consideration the Green's solution for the Poisson's equation,
the Eq.(\ref{gs}), it is deduced the relation:

\begin{equation}
\mu =\frac{\partial }{\partial \rho _{s}}s\left( \varepsilon _{s},\rho
_{s};\phi _{s}\right) -\beta m\phi _{s}-{\cal L}\left( {\bf r},-\frac{%
\partial }{\partial \phi _{s}}s\left( \varepsilon _{s},\rho _{s};\phi
_{s}\right) \right) \text{,}
\end{equation}
which is the same obtained as using the first Hamiltonian, the Eq.(\ref{HF}%
). Let us introduce the function $C\left( {\bf r}\right) $ as follows:

\begin{equation}
C\left( {\bf r}\right) =-{\cal L}\left( {\bf r},-\frac{\partial }{\partial
\phi _{s}}s\left( \varepsilon _{s},\rho _{s};\phi _{s}\right) \right) \text{.%
}  \label{cgs}
\end{equation}
Summarizing: the equations that dictate the equilibrium of the system are
the following:

\begin{equation}
\beta =\frac{\partial }{\partial \varepsilon _{s}}s\left( \varepsilon
_{s},\rho _{s};\phi _{s}\right) ,\text{ }\mu =\frac{\partial }{\partial \rho
_{s}}s\left( \varepsilon _{s},\rho _{s};\phi \right) -\beta m\phi _{s}+C%
\text{,}  \label{streq}
\end{equation}

\begin{equation}
\Delta \phi _{s}=4\pi Gm\rho _{s},\text{ \ \ }\Delta C=4\pi G\frac{\partial 
}{\partial \phi _{s}}s\left( \varepsilon _{s},\rho _{s};\phi _{s}\right) .
\label{es2}
\end{equation}
which have to satisfy the minimum request: the non-negativity of the matrix
functional of the Eq.(\ref{mf}). It is also necessary the exigency of the
conservation of the number of particles:

\begin{equation}
N=\int d^{3}{\bf r}\rho _{s}\left( {\bf r}\right) .
\end{equation}

The Planck's potential is expressed in terms of the local description as
follows:

${\cal P}\left( \beta _{o},N\right) =$%
\begin{equation}
\int d^{3}{\bf r}\ \beta {\cal H}\left[ \varepsilon _{s}\left( {\bf r}%
\right) ,\rho _{s}\left( {\bf r}\right) ,\phi _{s}\left( {\bf r}\right) %
\right] -s\left[ \varepsilon _{s}\left( {\bf r}\right) ,\rho _{s}\left( {\bf %
r}\right) ;\phi _{s}\left( {\bf r}\right) \right] \text{,}
\end{equation}
where $\beta $ is related with $\beta _{o}$ by the first relation of the Eq.(%
\ref{cer}). Using the Eq.(\ref{ah}), the above relation is rewritten
introducing the function $p\left[ \beta ,\rho ;\phi \right] $: 
\begin{equation}
{\cal P}\left( \beta _{o},N\right) =\int d^{3}{\bf r}\text{ }\beta \frac{1}{2%
}m\rho _{s}\left( {\bf r}\right) \phi _{s}\left( {\bf r}\right) +p\left[
\beta ,\rho _{s}\left( {\bf r}\right) ;\phi _{s}\left( {\bf r}\right) \right]
,  \label{plans}
\end{equation}
which is the Planck's potential density at the point ${\bf r}$. Due to the
scaling laws, it is convenient to set $N=1$, and let the dependency on $N$
to the scaling parameter $\alpha $. That is to consider $N$ as scaling
parameter:

\begin{equation}
\alpha =N\text{,}
\end{equation}
and set $N=1$ in all the above relations. In this case it will be performed
a scaling invariant description of the system. The equation system, the Eq.(%
\ref{es2}), must be solved under the constrain:

\begin{equation}
\int d^{3}{\bf r}\text{ }\rho _{s}\left( {\bf r}\right) =1.  \label{cons1}
\end{equation}

Finally, the energy scaling invariant is given by:

\begin{equation}
\epsilon \left( \beta \right) =\int d^{3}{\bf r}\text{ }\varepsilon
_{s}\left( {\bf r}\right) +\frac{1}{2}m\rho _{s}\left( {\bf r}\right) \phi
_{s}\left( {\bf r}\right) .  \label{enesi}
\end{equation}
In this way, when $N$ dependency is taken into account in the scaling laws,
the Eq.(\ref{els}),\ it is obtained a suitable problem that could be solved
by means of numerical computation. The microscopic model for the fluid must
be specified in order to continue the calculation. As example of
application, let us to apply this formalism in the next section to the
analysis of the classical system of identical non-interacting particles (at
the microscopic level).

As it can be seen, the consideration of the self-similarity scaling
postulates do not altere the description in the scaling invariant
equilibrium mean field theory ($N=1$). However, if the scaling laws are not
correctly chosen, this fact must lead to some unphysical consequences, such
as the non-proportionality of the Boltzmann's entropy and the Planck's
potential with the particle number, similarly to the Gibbs' paradox, and
therefore, it there will be a trivial ensemble inequivalence.

\section{Ideal Gas of Particles. Renormalization.}

The main difficulty in the statistical description of the astrophysical
systems is the existence of both, a short-range and long-range singularities
due to the consideration of the Newtonian gravitational interaction. The
first situation is the very-well known {\em gravothermal catastrophe} of the 
{\em N}-body self-gravitating system \cite{antonov,lbw}. In such a system,
there is no upper bound on the entropy and $\ $a state of arbitrarily large
entropy can be constructed from a centrally concentrated density profile by
shifting more of the mass towards the center (core-halo structures). It can
be seen in \cite{chava2,chava3,chava4} for review. This situation can be
easily avoided, since a new Physics appears at microscopic scales, i.e., the
Quantum Physics, which constitutes a natural renormalization when the system
is constituted by micro-particles: molecules, atoms and subatomic particles,
or in general way,\ by the consideration of the particles size.

The second, the long-range singularity, has a different nature. It is very
well-known that the gravitation is not able to confine the particles: it is
always possible that some of them have the sufficient energy for escaping
out from the system, so that, the system always undergoes an evaporation
process. Therefore, the astrophysical systems will never be in thermodynamic
equilibrium. However, there are intermediate stages where this behavior
might be neglected and a quasi--equilibrium state might be reached
(dynamical issues like ergodicity, mixing or ``approach to equilibrium''~ 
\cite{sasl,yawn,reidl}). In principle, these quasi-stationary stages can be
described dismissing the system evaporation. This last process could be
considered as a secondary effect, which modifies the quasi-stationary
equilibrium of the system. In the present approach, it will be only analyzed
the quasi-stationary equilibrium.

Let $K$ be the kinetic energy of a given particle, and $\phi \left( {\bf r}%
\right) $ its correspondent Newtonian potential energy at the point ${\bf r}$%
. That particle will be retained by the system gravity if the following
condition is hold:

\begin{equation}
K+m\phi <0\text{.}  \label{rc}
\end{equation}
When the above condition is not satisfied, the particle will be able to
scape out from the system if it does not lose its excessive energy. It will
be only consider in our description {\em those quasi-stationary stages in
which the above condition is hold for each particle of the system}. This
assumption is the key of the so called Michie-King models (see refs.\cite
{bin,inag,st,hjor,ing}). This exigency acts as a regularization procedure,
since it is sufficient to confine the system, avoiding in this way the
long-range singularity. No rigid boundaries are necessary in this case, so
that, no artificial parameters like the box volume are introduced \cite
{antonov,lbw}.

The local entropy density, that is, the function $s_{0}\left( \varepsilon
,\rho ;\phi \right) $, is obtained from the following model: Let us consider
a classical system of $N$ non-interacting particles which is confined by
means of a rigid boundary with volume $V$. Taking into consideration the
above renormalization prescription, the Eq.(\ref{rc}), the admissible stages
of this system are those in which the kinetic energy of each particle
satisfies the condition:

\begin{equation}
\frac{1}{2m}{\bf p}^{2}<U\text{,}
\end{equation}
where $m$ is the particle mass and $U$, the energy threshold. The
renormalized partition function of the canonical ensemble, $Z_{{\cal R}%
}\left( \lambda ,N;U\right) $, is given by:

$Z_{{\cal R}}\left( \lambda ,N;U\right) =$%
\begin{equation}
\mathrel{\mathop{\int }\limits_{\frac{1}{2m}{\bf p}_{k}^{2}<U,\text{ }k=1,..N}}%
\frac{1}{N!}\frac{d^{3N}Qd^{3N}P}{\left( 2\pi \hslash \right) ^{3N}}\exp
\left( -\lambda \stackrel{N}{%
\mathrel{\mathop{\sum }\limits_{k=1}}%
}\frac{1}{2m}{\bf p}_{k}^{2}\right) \text{,}
\end{equation}
where $\lambda $ is the canonical parameter. The calculation yields:

\begin{equation}
Z_{{\cal R}}\left( \lambda ,N;U\right) =\frac{V^{N}}{N!}\left( \frac{2m\pi }{%
\hbar ^{2}\lambda }\right) ^{\frac{3}{2}N}\left\{ F\left[ \left( \lambda
U\right) ^{\frac{1}{2}}\right] \right\} ^{N}\text{.}
\end{equation}
The function $F\left( z\right) $ in the above expression is defined by:

\begin{equation}
F\left( z\right) =\frac{4}{\sqrt{\pi }}\stackrel{z}{%
\mathrel{\mathop{\int }\limits_{0}}%
}x^{2}\exp \left( -x^{2}\right) dx,
\end{equation}
that is:

\begin{equation}
F\left( z\right) =%
\mathop{\rm erf}%
\left( z\right) -\frac{2}{\sqrt{\pi }}z\exp \left( -z^{2}\right) \text{.}
\end{equation}

In the FIG. 1., it is shown the behavior of this function. The asymptotic
dependency for low values of its argument is given by:

\begin{equation}
F\left( z\right) =\frac{4}{3\sqrt{\pi }}z^{3}-\frac{4}{5\sqrt{\pi }}%
z^{5}+O\left( z^{7}\right) \text{, for }z\lesssim 0.5\text{.}  \label{serie}
\end{equation}
The Planck's potential in the thermodynamic limit is given by:

$P\left( \lambda ,N;V,U\right) =$%
\begin{equation}
N\ln \left( \frac{N}{V}\right) -\frac{3}{2}N\ln \left( \frac{2m\pi }{\hbar
^{2}\lambda }\right) -N\ln F\left[ \left( \lambda U\right) ^{\frac{1}{2}}%
\right] ,
\end{equation}
and therefore, the energy is:

\begin{equation}
E=\frac{3N}{2\lambda }\left( 1-\frac{1}{3}\left. z\partial _{z}\ln F\left(
z\right) \right| _{z=\left( \lambda U\right) ^{\frac{1}{2}}}\right) \text{.}
\end{equation}
The caloric curve is shown in the FIG. 2. It can be seen the linear behavior
of the energy, at low values of the '{\em temperature}', $T=\lambda ^{-1}$,
which corresponds with the usual Maxwell's distribution. With the increasing
of the parameter $T$, the divergency between the renormalized model with the
ideal gas system becomes evident. This asymptotic dependency characterizes a
uniform distribution function for the particles momenta. In the FIG. 3., it
is shown the behavior of the distribution function of $p$ for different
values of the parameter $z=\left( \lambda U\right) ^{\frac{1}{2}}$. This
graphic shows the transition from a gaussian distribution for high values of 
$z$, to the uniform distribution at value $z=0$. Finally, the entropy
function is obtained by means of the Legendre's transformation:

\begin{equation}
S(E,N;V,U)=\lambda E-P\left( \lambda ,N;V,U\right) .
\end{equation}

\begin{figure}[tbp]
\caption{Behavior of the function $F\left( z\right) $. With the increasing
of $z$ this function tends fastly to the unity.}
\end{figure}

\begin{figure}[tbp]
\caption{Caloric curve of the microscopic model. At low ``temperatures'', $T=%
\protect\lambda ^{-1}$, the model behaves like the ordinary ideal gas, with
a maxwellian velocity distribution function, but at high temperatures, the
energy cutoff acts becoming homogeneous its velocity distribution function. }
\end{figure}

\begin{figure}[tbp]
\caption{Velocity distribution function at different values of $z$. All
these functions were normalized as the unity at the origen.}
\end{figure}

Summarizing: as consequence of the renormalization procedure assumed in the
Eq.(\ref{rc}), the local functions characterizing the local extensive
subsystem are given by:

\begin{equation}
p_{0}\left( \lambda ,\rho ,U\right) =\rho \ln \rho -\frac{3}{2}\rho \ln
\left( \frac{2m\pi }{\hbar ^{2}\lambda }\right) -\rho \ln F\left[ \left(
\lambda U\right) ^{\frac{1}{2}}\right] ,  \label{ppm}
\end{equation}

\begin{equation}
\varepsilon =\frac{3\rho }{2\lambda }\left( 1-\frac{1}{3}\left. z\partial
_{z}\ln F\left( z\right) \right| _{z=\left( \lambda U\right) ^{\frac{1}{2}%
}}\right) \text{,}
\end{equation}

\begin{equation}
s_{0}\left( \varepsilon ,\rho ;U\right) =\lambda \varepsilon -p_{0}\left(
\lambda ,\rho ,U\right) .
\end{equation}

It is not difficult to predict some of the consequences of the above
microscopic model. The energy threshold, $U$, is related with the Newtonian
potential as follows:

\begin{equation}
U=-m\phi \left( {\bf r}\right) \text{,}
\end{equation}
and therefore, the value of $U$ decreases from the inner regions of the
system to the outer ones. In the inner regions the local subsystems exhibit
the greater values of the parameter $z$, and therefore, their local
distribution functions for the particles momenta are almost gaussian.
However, at the outer regions, the local distribution functions diverge from
the gaussian shape, becoming asymptotically in a uniform distribution. It is
seen that the gravity drives the behavior of the local distribution function.

\section{Structure equations and system scaling laws.}

From the first equilibrium relation of the Eqs.(\ref{streq}) it is deduced
that the parameter $\lambda $ in the renormalized model is constant along
all the system and it is equal to $\beta $. The second equilibrium relation
allows us to obtain the particle distribution, that is, the particle density:

\begin{eqnarray*}
\mu &=&-\ln \rho -1+\frac{3}{2}\ln \left( \frac{2m\pi }{\hbar ^{2}\beta }%
\right) - \\
&&\text{ \ \ \ \ \ \ \ \ \ \ \ \ \ \ }-\ln F\left[ \left( -\beta m\phi
\right) ^{\frac{1}{2}}\right] -\beta m\phi +C,
\end{eqnarray*}
that is:

\begin{equation}
\rho =N\left( \beta ,\mu \right) \exp \left( -\beta m\phi +C\right) F\left[
\left( -\beta m\phi \right) ^{\frac{1}{2}}\right] \text{,}  \label{den}
\end{equation}
where the normalization constant $N\left( \beta ,\mu \right) $ is given by:

\begin{equation}
N\left( \beta ,\mu \right) =\left( \frac{2m\pi }{\hbar ^{2}\beta }\right) ^{%
\frac{3}{2}}\exp \left( -\mu -1\right) ,  \label{nc}
\end{equation}
(It will be obviated the subindex $s$ in the mean field description).

This last result allows us to specify the scaling parameter $\varkappa $ of
the system scaling laws, Eqs.(\ref{els}) and Eqs.(\ref{elp}), since from the
Eq.(\ref{nc}) the following relation is valid:

\begin{equation}
\varkappa =-\frac{3}{2}\pi =\frac{\varkappa +2}{2},
\end{equation}
and therefore:

\begin{equation}
\varkappa =2.  \label{kkk}
\end{equation}
Taking into consideration the relations of the Eq.(\ref{els}), the scaling
exponent constants are given by:

\[
\begin{tabular}{lll}
$m=2,$ & $\eta =\frac{4}{3},$ & $\pi =-\frac{4}{3},$%
\end{tabular}
\]

\begin{equation}
\begin{tabular}{ll}
$c=-\frac{1}{3},$ & $\chi =2,$%
\end{tabular}
\label{lenn}
\end{equation}
and therefore, the energy scaling exponent $\tau $ in the Eq.(\ref{esr}) is:

\begin{equation}
\tau =\frac{7}{3}.  \label{eel}
\end{equation}

This result differs from the proposed by Vega \& Sanchez in the refs.\cite
{veg1,veg2}. In these papers they studied \ the present model but this time
considering a box renormalization. They claimed correctly that the
thermodynamic limit for the astrophysical system must be invoked differently
from the usual extensive systems. In the Eq.(1) of the ref. \cite{veg1} they
demand that the thermodynamic limit is performed when:

\begin{equation}
N\rightarrow \infty ,\text{ }E\rightarrow \infty ,\text{ }L\rightarrow
\infty ,\frac{E}{N}=\text{{\em const}},\text{ }\frac{N}{L}=\text{{\em const},%
}
\end{equation}
where $L$ is the characteristic dimension of the box. However, it is easy to
see that this consideration leads to the non-proportionality of the
Boltzmann's entropy with the number of particles (see Eq.(8) or the
expression before the Eq.(12) in the ref.\cite{veg1}), and therefore,
ensemble inequivalence. In spite of this error, all the results obtained by
them are still correct, since they performed correctly the scaling invariant
description with the introduction of the {\em N}-independent parameters $\xi 
$ and $\eta $ defined as follows:

\begin{eqnarray}
\xi &=&\frac{EL}{Gm^{2}N^{2}}\text{ (in the microcanonical ensemble),} \\
\eta &=&\beta \frac{Gm^{2}N}{L}\text{ (in the canonical ensemble),}
\end{eqnarray}
which are also in agreement with the scaling laws considered by us in the
Eqs.(\ref{elp}),\ (\ref{esr}), (\ref{lenn}), and (\ref{eel}). In their
approach, they assumed incorrectly that the mechanical energy scales
proportional to $N$. Our energy scaling law suggests that this model can not
be applied for a number arbitrarily large of particles: {\em an upper bound
for }$N${\em \ appears} for the validity of the non relativistic conditions
(see in section VI).

As it can be easily seen, in the Eq.(\ref{den}), when $\phi \rightarrow 0$,
the density vanishes, $\rho \rightarrow 0$. Thus, our model is renormalized.
In order to obtain the structure equations, the Eqs.(\ref{es2}), it is
necessary to perform the following calculation:

\begin{eqnarray}
-\frac{\partial }{\partial \phi }s_{0}\left( \varepsilon ,\rho ;\phi \right)
&=&\frac{\partial }{\partial \phi }p_{0}\left( \beta ,\rho ;\phi \right) 
\nonumber \\
&=&\beta m\rho \left. \frac{1}{2z}\frac{\partial }{\partial z}\ln F\left(
z\right) \right| _{z=\left( -\beta m\phi \right) ^{\frac{1}{2}}}.
\end{eqnarray}

Introducing the constant $K$:

\begin{equation}
K=Gm^{2}\beta N\left( \beta ,\mu \right) ,  \label{kde}
\end{equation}
the dimensionless function $\Phi $:

\begin{equation}
\Phi =-\beta m\phi ,
\end{equation}
and the coordinate $\xi $:

\begin{equation}
\xi =K^{\frac{1}{2}}r,
\end{equation}
the structure equations are finally written as follows:

\begin{equation}
\Delta \Phi =-4\pi F_{1}\left( C,\Phi \right) ,\text{ \ \ \ }\Delta C=-4\pi
F_{2}\left( C,\Phi \right) \text{\ ,}  \label{ple}
\end{equation}
where:

\begin{eqnarray}
F_{1}\left( C,\Phi \right) &=&\exp \left( C+\Phi \right) F\left( \Phi ^{%
\frac{1}{2}}\right) ,  \nonumber \\
F_{2}\left( C,\Phi \right) &=&\frac{2}{\sqrt{\pi }}\exp \left( C\right) \Phi
^{\frac{1}{2}}.
\end{eqnarray}

Due to the scaling transformation, the constant $K$ depends on $N$ as
follows:

\begin{equation}
K=N^{\frac{2}{3}}K_{o},
\end{equation}
where $K_{o}$ is obtained substituting $\beta $ by $\beta _{o}$ in the Eq.(%
\ref{kde}). The coordinate $\xi $ is {\em N}-independent.

\section{Numerical Analysis}

In order to perform the numerical analysis of the above equations, it is
convenient to introduce the following constant: 
\begin{equation}
\epsilon _{0}=2\pi \frac{G^{2}m^{5}}{\hbar ^{2}},\text{ }\rho _{0}=\left(
2\pi \right) ^{3}\frac{G^{3}m^{9}}{\hbar ^{6}},\text{ \ }l_{0}=\allowbreak 
\frac{1}{2\pi }\frac{\hbar ^{2}}{Gm^{3}},  \label{units}
\end{equation}
which are respectively the energy, density, and length characteristic units
of the present analysis. The above consideration allows us to set $\hbar
=G=m=1$. Hearafter it will be imposed the condition $N=1$.

In terms of the functions $C$ and $\Phi $, the particle number constrain is
written as follows: 
\begin{equation}
\exp (-\mu -1)\frac{K_{o}^{\frac{3}{2}}}{\beta _{o}^{\frac{3}{2}}}\stackrel{%
+\infty }{%
\mathrel{\mathop{\int }\limits_{0}}%
}4\pi \xi ^{2}d\xi F_{1}\left( C,\Phi \right) =1\text{,}  \label{ncc}
\end{equation}
which establishes a functional relation between the thermodynamic parameters 
$\beta _{o}$ and $\mu $. This exigency can be rewritten analyzing the
behavior of the functions $C$ and $\Phi $ in the asymptotic region $\xi
\rightarrow +\infty $. The spherical solution of the Poisson's equation
satisfy the following asymptotic relation:

\begin{equation}
\mathrel{\mathop{\lim }\limits_{r\rightarrow \infty }}%
r^{2}\nabla _{r}\phi \left( r\right) =GM\text{,}
\end{equation}
where $M$ is the total mass of the system. In the present problem, taking
into consideration the characteristic units, the Eq.(\ref{units}), the above
relation is rewritten as follows:

\begin{equation}
\mathrel{\mathop{\lim }\limits_{\xi \rightarrow \infty }}%
\xi ^{2}\frac{\partial }{\partial \xi }\Phi \left( \xi \right) =-\beta
_{o}K_{o}^{\frac{1}{2}}.  \label{bcs}
\end{equation}

In the asymptotic region $\xi \rightarrow \infty $, the functions $\Phi $
and $C$ vanish identically:

\begin{equation}
\mathrel{\mathop{\lim }\limits_{\xi \rightarrow \infty }}%
\Phi \left( \xi \right) =%
\mathrel{\mathop{\lim }\limits_{\xi \rightarrow \infty }}%
C\left( \xi \right) =0\text{,}  \label{bcc}
\end{equation}
due to the asymptotic behavior of the Green's function, Eq.(\ref{tgf}), in
their representations, Eq.(\ref{gs}) and Eq.(\ref{cgs}). The numerical
solution of the Poisson's like equations system, Eq.(\ref{ple}), is carried
out imposing the boundary at the origin $\xi =0$. This is performed
demanding at the origin:

\begin{eqnarray}
\Phi \left( 0\right) &=&\Phi _{0}>0\text{,} \\
C\left( 0\right) &=&C_{0.}
\end{eqnarray}

\ The value of $C_{0}$ must be selected appropriately since it must vanish
when $\xi \rightarrow \infty $, and therefore, $C_{0}$ depends on the
parameter $\Phi _{0}$. This situation can be overcome redefining the problem
as follows: firstly, displacing the function $C\left( \xi \right) $:

\begin{equation}
C\left( \xi \right) =-c\left( \infty \right) +c\left( \xi ^{^{\prime
}}\right) ,
\end{equation}
where $c\left( \xi ^{^{\prime }}\right) $ is the solution of the Poisson's
like equation system with the following boundary conditions:

\begin{eqnarray}
\Phi \left( 0\right) &=&\Phi _{0}\text{,} \\
c\left( 0\right) &=&0,
\end{eqnarray}
and the new coordinate $\xi ^{^{\prime }}$ is related with $\xi $ throughout
the relation:

\begin{equation}
\xi ^{^{\prime }}=\exp \left[ -\frac{1}{2}c\left( \infty \right) \right] \xi
.
\end{equation}

Hereafter it is obviated the punctuation in $\xi $. The Eq.(\ref{bcs}) is
rewritten as follows:

\begin{equation}
h\left( \Phi _{0}\right) \exp \left[ \frac{1}{2}c\left( \infty \right) %
\right] =\beta _{o}K_{o}^{\frac{1}{2}}=\beta _{o}^{\frac{3}{4}}\exp \left[ -%
\frac{1}{2}\left( \mu +1\right) \right] .  \label{hdf}
\end{equation}
where the function $h\left( \Phi _{0}\right) $ was introduced:

\begin{equation}
h\left( \Phi _{0}\right) =-%
\mathrel{\mathop{\lim }\limits_{\xi \rightarrow \infty }}%
\xi ^{2}\frac{\partial }{\partial \xi }\Phi \left( \xi \right) \text{.}
\label{hd}
\end{equation}

So far, there are three parameters involved in the solution of the problem, $%
\beta _{o}$, $\mu $ and $\Phi _{0}$. The above relation determines $\mu $ as
a function of $\beta _{o}$ and $\Phi _{0}$. An additional relation is needed
to specify the functional dependency between $\beta _{o}$ and $\Phi _{0}$.
The solution to this situation is obtained throughout the equivalency
between the statistical ensembles when {\em this equivalency held}. From the
Legendre's transformation is followed that the energy is obtained from the
Planck's potential through the relation:

\begin{equation}
\epsilon \left( \beta _{o},\Phi _{0}\right) =\frac{d}{d\beta _{o}}{\cal P}%
\left( \beta _{o},\Phi _{0}\right) ,
\end{equation}
and therefore:

\begin{equation}
\frac{d\beta _{o}\left( \Phi _{0}\right) }{d\Phi _{0}}=\frac{\partial }{%
\partial \Phi _{0}}{\cal P}\left( \beta _{o},\Phi _{0}\right) \left[
\epsilon \left( \beta _{o},\Phi _{0}\right) -\frac{\partial }{\partial \beta
_{o}}{\cal P}\left( \beta _{o},\Phi _{0}\right) \right] ^{-1}.  \label{sequa}
\end{equation}
This is the second relation that we have been looking for.

From the Eq.(\ref{hdf}) it is deduced the functional dependency between the
chemical potential $\mu $ and the parameters $\beta _{o}$ and $\Phi _{0}$:

\begin{equation}
\mu =-1+\frac{3}{2}\ln \beta _{o}-2\ln h\left( \Phi _{0}\right) -c\left(
\infty \right) .
\end{equation}

Using the Eq.(\ref{enesi}), the energy is expressed by:

\begin{equation}
\epsilon \left( \beta _{o},\Phi _{0}\right) =\frac{1}{\beta _{o}}\left[ 
\frac{3}{2}-\frac{h_{1}\left( \Phi _{0}\right) }{h\left( \Phi _{0}\right) }%
\right] ,
\end{equation}
where $h_{1}\left( \Phi _{0}\right) $ is defined as:

\begin{equation}
h_{1}\left( \Phi _{0}\right) =\stackrel{+\infty }{%
\mathrel{\mathop{\int }\limits_{0}}%
}4\pi \xi ^{2}d\xi \Phi \left[ F_{2}\left( c,\Phi \right) +\frac{1}{2}%
F_{1}\left( c,\Phi \right) \right] .
\end{equation}

Similarly, the Planck's potential in the Eq.(\ref{plans}) is rewritten as
follows:

\begin{equation}
{\cal P}\left( \beta _{o},\Phi _{0}\right) =-\frac{3}{2}\ln \beta _{o}+2\ln
h\left( \Phi _{0}\right) +\frac{h_{2}\left( \Phi _{0}\right) }{h\left( \Phi
_{0}\right) },
\end{equation}
where $h_{2}\left( \Phi _{0}\right) $ is defined by the expression:

\begin{equation}
h_{2}\left( \Phi _{0}\right) =\stackrel{+\infty }{%
\mathrel{\mathop{\int }\limits_{0}}%
}4\pi \xi ^{2}d\xi F_{1}\left( c,\Phi \right) \left[ c+\frac{1}{2}\Phi %
\right] .
\end{equation}

Introducing the function $H\left( \Phi _{0}\right) $:

\begin{equation}
H\left( \Phi _{0}\right) =2\ln h\left( \Phi _{0}\right) +\frac{h_{2}\left(
\Phi _{0}\right) }{h\left( \Phi _{0}\right) },
\end{equation}
the Eq.(\ref{sequa}) is rewritten as follows:

\begin{equation}
\frac{d\ln \beta _{o}\left( \Phi _{0}\right) }{d\Phi _{0}}=\frac{\partial
H\left( \Phi _{0}\right) }{\partial \Phi _{0}}\left[ 3-\frac{h_{1}\left(
\Phi _{0}\right) }{h\left( \Phi _{0}\right) }\right] ^{-1}.  \label{betequa}
\end{equation}
Finally, the entropy of the system is given by:

\begin{equation}
S\left[ \epsilon \left( \beta _{o},\Phi _{0}\right) \right] =\frac{3}{2}+%
\frac{3}{2}\ln \beta _{o}-H\left( \Phi _{0}\right) -\frac{h_{1}\left( \Phi
_{0}\right) }{h\left( \Phi _{0}\right) }.
\end{equation}

Of course, the validity of the Eq.(\ref{betequa}) rests on the equivalency
between the micro and canonical ensembles, which is satisfied wherever the
parameter $\beta _{o}$ is a decreasing function of the energy:

\begin{equation}
\frac{d\beta _{o}\left( \Phi _{0}\right) }{d\epsilon \left( \Phi _{0}\right) 
}<0.  \label{conequi}
\end{equation}
Introducing the function ${\cal R}\left( \Phi _{0}\right) $:

\begin{equation}
{\cal R}\left( \Phi _{0}\right) =\stackrel{\Phi _{0}}{%
\mathrel{\mathop{\int }\limits_{0}}%
}\frac{\partial H\left( s\right) }{\partial s}\left[ 3-\frac{h_{1}\left(
s\right) }{h\left( s\right) }\right] ^{-1}ds,
\end{equation}
the general solution of the Eq.(\ref{betequa}) is expressed as follows:

\begin{equation}
\beta _{o}\left( \Phi _{0}\right) =C\exp \left[ {\cal R}\left( \Phi
_{0}\right) \right] ,
\end{equation}
where the integration constant $C$ could be fixed as the unity without loss
of generality. Using the function ${\cal R}\left( \Phi _{0}\right) $ and the
function ${\cal K}\left( \Phi _{0}\right) $:

\begin{equation}
{\cal K}\left( \Phi _{0}\right) =\frac{3}{2}-\frac{h_{1}\left( \Phi
_{0}\right) }{h\left( \Phi _{0}\right) },
\end{equation}
the condition of the Eq.(\ref{conequi}) is rewritten as:

\begin{equation}
\frac{d}{d\Phi _{0}}{\cal K}\left( \Phi _{0}\right) /\frac{d}{d\Phi _{0}}%
{\cal R}\left( \Phi _{0}\right) -{\cal K}\left( \Phi _{0}\right) <0.
\end{equation}

The Eq.(\ref{betequa}) will be justified when the above relation is
satisfied, otherwise, {\em at first sight}, the results obtained from this
methodology has apparently no sense. According to the results of the ref. 
\cite{vel2}, the above observation is correct: {\em in the representation} $%
\left( {\cal E},N\right) $ the equivalency of the ensemble demand the
validity of the condition given in the Eq.(\ref{conequi}). Nevertheless, 
{\em it can be chosen another representation for the integrals of motion in
which the equivalence between the ensemble takes place}. For example, the
canonical description could be performed introducing the following
representation for the integrals of motion:

\[
\left( {\cal E},N\right) \rightarrow \left( {\cal U}_{\varphi },N\right) 
\text{, where }{\cal U}_{\varphi }=N\varphi \left( {\cal E}/N\right) \text{,}
\]
where $\varphi $ is an arbitrary piecewise monotonic function of $\epsilon =%
{\cal E}/N$ at least two times differentiable with exception of the juncture
boundary. As it can be easily seen, ${\cal U}_{\varphi }$ and ${\cal E}$
possess the {\em same} scaling law in the thermodynamic limit $N\rightarrow
\infty $, that is the reason why both representations are adequate in the
generalized canonical ensemble with probabilistic distribution function
given by:

\begin{equation}
\omega _{c}\left( {\cal U}_{\varphi };\beta _{\varphi },N\right) =\frac{1}{%
Z_{\varphi }\left( \beta _{\varphi },N\right) }\exp \left( -\beta _{\varphi
}\cdot {\cal U}_{\varphi }\right) ,
\end{equation}
where $Z_{\varphi }\left( \beta _{\varphi },N\right) $ is the partition
function in the present representation from which it is derived the
corresponding Planck's potential:

\begin{equation}
{\cal P}_{\varphi }\left( \beta _{\varphi },N\right) =-\ln Z_{\varphi
}\left( \beta _{\varphi },N\right) .
\end{equation}
It is easy to see that in this case it is also valid the Legendre's
transformation between the thermodynamic potentials when the thermodynamic
limit is performed:

\begin{equation}
S\left( {\cal U}_{\varphi },N\right) \simeq \beta _{\varphi }\cdot {\cal U}%
_{\varphi }-{\cal P}_{\varphi }\left( \beta _{\varphi },N\right) ,
\end{equation}
with:

\begin{equation}
{\cal U}_{\varphi }=\frac{\partial }{\partial \beta _{\varphi }}{\cal P}%
_{\varphi }\left( \beta _{\varphi },N\right) .
\end{equation}

During the representation changes the canonical parameter and the Planck'
potential are transformed as follows:

\begin{eqnarray}
\beta _{o} &\rightarrow &\beta _{\varphi }=\left( \frac{d\varphi \left(
\epsilon \right) }{d\epsilon }\right) ^{-1}\beta _{o}, \\
{\cal P} &\rightarrow &{\cal P}_{\varphi }={\cal P}+\left( \frac{d\varphi
\left( \epsilon \right) }{d\epsilon }\right) ^{-1}\left( \varphi \left(
\epsilon \right) -\epsilon \frac{d\varphi \left( \epsilon \right) }{%
d\epsilon }\right) \beta _{o}N,
\end{eqnarray}
in the way that the entropy remains unchanged:

\begin{equation}
S\left( {\cal E},N\right) \rightarrow S\left( {\cal U}_{\varphi },N\right)
\equiv S\left( {\cal E},N\right) .
\end{equation}

In the new representation the equivalence of the microcanonical and
canonical ensembles takes place when it is satisfied the condition:

\begin{equation}
\frac{d^{2}}{d{\cal U}_{\varphi }^{2}}S\left( {\cal U}_{\varphi },N\right)
\equiv \beta _{\varphi }\frac{d}{d\varphi }\ln \beta _{\varphi }<0,
\end{equation}
that is, the negative definition of the curvature tensor asociate to the new
representation. This condition can be rewritten again as follows:

\begin{equation}
\left( \frac{d\varphi \left( \epsilon \right) }{d\epsilon }\right)
^{-2}\beta _{o}\frac{d}{d\epsilon }\ln \left[ \left( \frac{d\varphi \left(
\epsilon \right) }{d\epsilon }\right) ^{-1}\beta _{o}\right] <0.
\label{cnega}
\end{equation}
Let us analyze a possible way for the selection of the function\ $\varphi $.
The Eq.(\ref{cnega}) can be rewritten as:

\[
\left( \frac{d\varphi \left( \epsilon \right) }{d\epsilon }\right) ^{-2}%
\left[ \frac{d\beta _{o}}{d\epsilon }-\beta _{o}\frac{d}{d\epsilon }\ln
\left( \frac{d\varphi \left( \epsilon \right) }{d\epsilon }\right) \right] 
\]
\begin{equation}
\equiv -\left( \frac{d\varphi \left( \epsilon \right) }{d\epsilon }\right)
^{-2}a\left( \epsilon \right) ,
\end{equation}
where the function\ $a\left( \epsilon \right) >0$, from the which is derived:

\begin{equation}
\frac{d\varphi \left( \epsilon \right) }{d\epsilon }=C\beta _{o}\left(
\epsilon \right) \exp \left( \int \frac{a\left( \epsilon \right) }{\beta
_{o}\left( \epsilon \right) }d\epsilon \right) ,
\end{equation}
where $C$ is a positive constant which could be set as unity. It is easy to
see that a convenient choice for the function $a\left( \epsilon \right) $,
from the theoretical viewpoint, could be given by:

\begin{equation}
a\left( \epsilon \right) =\left\{ 
\begin{tabular}{lll}
$-k\left( \epsilon \right) $ & if & $k\left( \epsilon \right) <0,$ \\ 
$\beta _{o}^{2}\left( \epsilon \right) =\beta _{c}^{2}\sim const$ \ \  & if
& $k\left( \epsilon \right) =0,$ \\ 
$k\left( \epsilon \right) $ & if & $k\left( \epsilon \right) >0,$%
\end{tabular}
\right|
\end{equation}
where it was introduced the nomenclature:

\begin{equation}
k\left( \epsilon \right) =\frac{d\beta _{o}\left( \epsilon \right) }{%
d\epsilon }=\frac{d^{2}}{d\epsilon ^{2}}s_{B}\left( \epsilon \right) ,
\end{equation}
where $s_{B}\left( \epsilon \right) $ is the Boltzmann's entropy per
particle. In this representation change:

\begin{equation}
\frac{d\varphi \left( \epsilon \right) }{d\epsilon }=\left\{ 
\begin{tabular}{lll}
$1$ & if & $k\left( \epsilon \right) <0,$ \\ 
$\beta _{c}\exp \left( \beta _{c}\epsilon \right) $ & if & $k\left( \epsilon
\right) =0,$ \\ 
$\beta _{o}^{2}\left( \epsilon \right) $ & if & $k\left( \epsilon \right)
>0. $%
\end{tabular}
\right|
\end{equation}

As it can be seen, in this representation $\varphi \left( \epsilon \right) $
is an piecewise monotonic function of $\epsilon $, and therefore, there is a
bijective correspondence between $\epsilon $ and $\varphi \left( \epsilon
\right) $. In this way it have been shown that it is always possible to
chose a representation in which it is valid the correspondence among the
ensembles allowing us to extend the applicability of the generalized
canonical ensembles to those situations in which the\ ordinary heat capacity
is negative.

The reparametrization freedom of the microcanonical ensemble allows us to
extend the validity of the mean field approach with an appropriate selection
of the function $\varphi $. It can be easily shown that all the results of
the canonical description in the different representations are only
non-coincident when it is taken into account those observables involving
second derivatives (or superior) of the thermodynamic potentials. That is
the reason why it is considered that the Eq.(\ref{betequa}) can be extended
to those regions where there is an univocal dependence of the canonical
parameter $\beta _{o}$ with the energy $\epsilon $.

In order to solve the Eq.(\ref{betequa}) it is necessary to known the
asymptotic behavior of the functions $h\left( \Phi _{0}\right) $, $%
h_{1}\left( \Phi _{0}\right) $ and $h_{2}\left( \Phi _{0}\right) $ for small
values of $\Phi _{0}$. This is done developing the functions $c$ and $\Phi $
in a perturbative expansion of the parameter $\Phi _{0}$:

\begin{equation}
\Phi =\overline{\Phi }+\delta \Phi +...,\text{ }c=\overline{c}+\delta c+...
\end{equation}
as well as the functions $F_{1}\left( c,\Phi \right) $ and $F_{2}\left(
c,\Phi \right) $:

\begin{eqnarray}
F_{1}\left( c,\Phi \right) &=&\overline{F}_{1}\left( \overline{c},\overline{%
\Phi }\right) +\delta F_{1}\left( \delta c,\delta \Phi ;\overline{c},%
\overline{\Phi }\right) +...,  \nonumber \\
F_{2}\left( c,\Phi \right) &=&\overline{F}_{2}\left( \overline{c},\overline{%
\Phi }\right) +\delta F_{2}\left( \delta c,\delta \Phi ;\overline{c},%
\overline{\Phi }\right) +...,
\end{eqnarray}
where the problem given in the Eq.(\ref{ple}) is rewritten as follows:

\begin{equation}
\Delta \overline{\Phi }=-4\pi \overline{F}_{1}\left( \overline{c},\overline{%
\Phi }\right) ,\text{ \ \ \ }\Delta c=-4\pi \overline{F}_{1}\left( \overline{%
c},\overline{\Phi }\right) ,
\end{equation}

\begin{equation}
\Delta \left( \delta \Phi \right) =-4\pi \delta F_{1}\left( \delta c,\delta
\Phi ;\overline{c},\overline{\Phi }\right) ,
\end{equation}

\begin{equation}
\Delta \left( \delta c\right) =-4\pi \delta F_{2}\left( \delta c,\delta \Phi
;\overline{c},\overline{\Phi }\right) ,...
\end{equation}

Using the power expansion of the function $F\left( z\right) $, Eq.(\ref
{serie}), it is very easy to obtain the following relations:

\begin{equation}
\overline{F}_{1}\left( \overline{c},\overline{\Phi }\right) =\exp \left( 
\overline{c}\right) \frac{4}{3\sqrt{\pi }}\overline{\Phi }^{\frac{3}{2}},%
\text{ }\overline{F}_{2}\left( \overline{c},\overline{\Phi }\right) =\exp
\left( \overline{c}\right) \frac{2}{\sqrt{\pi }}\overline{\Phi }^{\frac{1}{2}%
},
\end{equation}

\begin{equation}
\delta F_{1}\left( \delta c,\delta \Phi ;\overline{c},\overline{\Phi }%
\right) =\exp \left( \overline{c}\right) \frac{2}{\sqrt{\pi }}\left[ -\frac{2%
}{5}\overline{\Phi }^{\frac{5}{2}}+\frac{2}{3}\overline{\Phi }^{\frac{3}{2}%
}\delta c+\overline{\Phi }^{\frac{1}{2}}\delta \Phi \right] ,
\end{equation}

\[
\delta F_{2}\left( \delta c,\delta \Phi ;\overline{c},\overline{\Phi }%
\right) =\exp \left( \overline{c}\right) \frac{2}{\sqrt{\pi }}\left[ 
\overline{\Phi }^{\frac{1}{2}}\delta c+\frac{1}{2}\overline{\Phi }^{-\frac{1%
}{2}}\delta \Phi \right] . 
\]

In this asymptotic region, the structure equations (the Eq.(\ref{ple}))
become in the following problem:

\begin{equation}
\Delta \overline{\Phi }=-4\pi \exp \left( \overline{c}\right) \frac{4}{3%
\sqrt{\pi }}\overline{\Phi }^{\frac{3}{2}},\text{ \ }\Delta \overline{c}%
=-4\pi \exp \left( \overline{c}\right) \frac{2}{\sqrt{\pi }}\overline{\Phi }%
^{\frac{1}{2}},
\end{equation}

\begin{equation}
\overline{\Phi }\left( 0\right) =\Phi _{0},\text{ }\overline{\Phi }^{\prime
}\left( 0\right) =0\text{ and }\overline{c}\left( 0\right) =0,\text{ }%
\overline{c}^{\prime }\left( 0\right) =0.
\end{equation}
It is very easy to see that the above problem possesses a fractal
characteristic and is quite similar to the polytropic model with polytropic
index $\gamma =\frac{5}{3}$, which allows us to express its general solution
as follows:

\begin{equation}
\overline{\Phi }=\Phi _{0}\varphi \left( \Phi _{0}^{\frac{1}{4}}\xi \right) ,%
\text{ }\overline{c}=\psi \left( \Phi _{0}^{\frac{1}{4}}\xi \right) ,
\end{equation}
where the functions $\varphi \left( z\right) $ and $\psi \left( z\right) $
are the solution of the problem:

\begin{equation}
\Delta _{z}\varphi =-4\pi \exp \left( \psi \right) \frac{4}{3\sqrt{\pi }}%
\varphi ^{\frac{3}{2}},\text{ }\Delta _{z}\psi =-4\pi \exp \left( \psi
\right) \frac{2}{\sqrt{\pi }}\varphi ^{\frac{1}{2}},
\end{equation}
with boundary conditions:

\begin{equation}
\varphi \left( 0\right) =1\text{ and }\varphi ^{\prime }\left( 0\right) =0,%
\text{ }\psi \left( 0\right) =0\text{ and }\psi ^{\prime }\left( 0\right) =0.
\end{equation}
\qquad

It is interesting to point out that this polytropic index characterizes an
adiabatic process of the ideal gas of particles, which in our case is the
evaporation of the system in the vacuum. However, this equation system is
not equivalent to polytropic model due to the presence of the term of the
gravity driving, the pres{}ence of the function $\overline{c}$. It is easy
to understand that the maximum effects of this term appear in the outer
region of the system, in the halo. In this region the results of the
polytropic model and the pseudo-polytropic model given by the Eq.(\ref{ppm})
are different. Moreover, this kind of behavior is also present for all the
values of the parameter $\Phi _{0}$, and as a consequence of the
renormalization prescription assumed in Eq.(\ref{rc}), it leads to
configurations of the systems characterized by an isothermal core with a
quasi-polytropic halo. The effect of the function $\overline{c}$ is showed
in the FIG.4. As it can be seen, the consideration of this term reduces the
system size in comparison with the polytropic equation.

\begin{figure}[tbp]
\caption{Comparison between the polytropic model with $\protect\gamma =\frac{%
5}{3}$ and the quasi-polytropic model presented in the present analysis.}
\end{figure}

Similarly, the functions $\delta c$ and $\delta \Phi $ are expressed as
follows:

\begin{equation}
\delta \Phi =\Phi _{0}^{2}\varphi _{1}\left( \Phi _{0}^{\frac{1}{4}}\xi
\right) ,\text{ }\delta c=\Phi _{0}\psi _{1}\left( \Phi _{0}^{\frac{1}{4}%
}\xi \right) ,
\end{equation}
where the functions $\varphi _{1}\left( z\right) $ and $\psi _{1}\left(
z\right) $ are obtained from the problem:

\begin{equation}
\Delta _{z}\varphi _{1}=-4\pi \exp \left( \psi \right) \frac{2}{\sqrt{\pi }}%
\left[ -\frac{2}{5}\varphi ^{\frac{5}{2}}+\frac{2}{3}\varphi ^{\frac{3}{2}%
}\psi _{1}+\varphi ^{\frac{1}{2}}\varphi _{1}\right] ,
\end{equation}

\begin{equation}
\Delta _{z}\psi _{1}=-4\pi \exp \left( \psi \right) \frac{2}{\sqrt{\pi }}%
\left[ \varphi ^{\frac{1}{2}}\psi _{1}+\frac{1}{2}\varphi ^{-\frac{1}{2}%
}\varphi _{1}\right] ,
\end{equation}
with boundary conditions:

\begin{equation}
\varphi _{1}\left( 0\right) =0\text{ and }\varphi _{1}^{\prime }\left(
0\right) =0,\text{ }\psi _{1}\left( 0\right) =0\text{ and }\psi _{1}^{\prime
}\left( 0\right) =0.
\end{equation}

In this way it is obtained the asymptotic dependence of the functions $%
h\left( s\right) $, $h_{1}\left( s\right) $ and $h_{2}\left( s\right) $:

\begin{eqnarray}
h\left( s\right) &=&a_{1}s^{\frac{3}{4}}+a_{2}s^{\frac{7}{4}}+O\left( s^{%
\frac{7}{4}}\right) ,\text{ }  \label{h00} \\
h_{1}\left( s\right) &=&b_{1}s^{\frac{3}{4}}+b_{2}s^{\frac{7}{4}}+O\left( s^{%
\frac{7}{4}}\right) ,  \label{h11} \\
h_{2}\left( s\right) &=&c_{1}s^{\frac{3}{4}}+c_{2}s^{\frac{7}{4}}+O\left( s^{%
\frac{7}{4}}\right) .  \label{h22}
\end{eqnarray}
The calculation yields:

\begin{equation}
\begin{tabular}{l}
$a_{1}=0.744,$ $a_{2}=-0.143,$ \\ 
$b_{1}=1.5a_{1}=\allowbreak 1.\,\allowbreak 116,$ $b_{2}=-0.242,$ \\ 
$c_{1}=-0.622,$ $c_{2}=0.478.$%
\end{tabular}
\end{equation}

Thus, in the asymptotic region, the relation between the canonical parameter 
$\beta _{o}$ and $\Phi _{0}$ is given by:

\begin{equation}
\beta _{o}\left( \Phi _{0}\right) =\Phi _{0}+q\Phi _{0}^{2}+\allowbreak
O\left( \Phi _{0}^{2}\right) ,
\end{equation}
where:

\begin{equation}
q\simeq 0.025.
\end{equation}
Finally, the energy in this limit is given by:

\begin{equation}
\epsilon \simeq 0.\,\allowbreak 055+O(\Phi _{0}).
\end{equation}

\section{Results and Discussions}

Let us begin the discussion analyzing the scaling laws of our system. In
general way, the scaling laws {\em determine} the specific form of the
thermodynamic formalism. The astrophysical systems exhibit exponential
self-similarity scaling laws in the thermodynamic limit, which is the reason
why the Boltzmann-Gibbs' Statistics is applicable to the macroscopic
description of this kind of systems in spite of that they are nonextensive.
This analysis can be performed choosing an adequate selection of the
representation of the integrals of motion. As it was shown in the section
IV, the {\em N}-body Newtonian self-gravitating gas exhibits self-similarity
properties under the following thermodynamic limit:

\begin{equation}
N\rightarrow \infty ,\text{ }E\rightarrow \infty ,\text{ }L\rightarrow 0,%
\text{ }\frac{E}{N^{\frac{7}{3}}}=const.,\text{ }LN^{\frac{1}{3}}=const.,
\end{equation}
where $L$ is a characteristic linear dimension of the system. This result
differed from the obtained by Vega \& Sanchez in refs.\cite{veg1,veg2}.

Many investigators do not pay so much attention to the analysis of the
system scaling laws when it is performed its macroscopic description.
Fortunately, most of the systems studied so far belong to the class of the
pseudoextensive systems. When it is performed the {\em N}-independent
description, the results of the analysis do not depend on the scaling laws,
and therefore, many of these results remain valid. However, the non or bad
consideration of the scaling laws leads in many case to the trivial {\em %
nonequivalence of the statistics ensembles}.

As example of the above affirmation it can be seen the anomalies presented
in the dynamical study of the self-gravitating systems performed by
Cerruti-Sola \& Pettini in the ref.\cite{pet}. In that paper, they observed
a {\em weakening} of the system chaotic behavior with the increasing of the
particles number $N$. It is very well-known the consequence of this fact on
the ergodicity of the system: the chaotic dynamics provides the mixing
property in the phase space necessary for obtaining the equilibrium. In that
example the chaoticity time grows with $N$, and therefore, the systems could
expend so much time to arrive to the equilibrium. Similar behavior has been
seen in the dynamical study of the so called Hamiltonian Mean Field model
(see for example in the ref.\cite{lato}).

However, our analysis allows\ us to understand the origin of this behavior,
a least, for the self-gravitating gas. In that study it was considered that
the energy is scaled proportional to $N$ during the realization
thermodynamic limit, which is a false assumption. It is very easy to see
that the dependency of the instability exponent $\lambda _{H}$ of the energy
(Fig.4. of the ref.\cite{pet}) {\em is corrected} when it is considered the
right scaling law of the energy: its proportionality to $N^{\frac{7}{3}}$.
The anomalies disappear when it is assumed this scaling law for the energy.
Anyway, in spite of they assumed a wrong scaling law for the energy, they
obtained the correct dependency of the instability exponent $\lambda _{H}$
with the energy per particle: the power law $\epsilon ^{\frac{3}{2}}$. Our
analysis suggests that the anomalies presented in the dynamical study of the
Hamiltonian Mean Field model {\em could possess the same origin}. In future
works we will consider this possibility .

Some other consequences of this scaling laws are found analyzing the limit
of applicability of the model. In the present work it was considered a
classical gas of identical particles with mass $m$. Taking into account the
rest energy of the particles, the nonrelativistic limit is valid when the
absolute value of the mechanical energy of the system is much smaller than
its rest energy :

\begin{equation}
\left| \epsilon _{0}\epsilon \left( \Phi _{0}\right) N^{\frac{7}{3}}\right|
\ll mc^{2}N,
\end{equation}
where $\epsilon _{0}$ is the characteristic energy of the model, Eq.(\ref
{units}). However, this condition can not be satisfied for an arbitrary
number of particles. In fact, when $N$\ tends to $N_{0}$:

\begin{equation}
N_{0}=\left( 2\pi \frac{mc^{2}}{\epsilon _{0}}\right) ^{\frac{3}{4}}=\left( 
\frac{\hbar c}{G}\right) ^{\frac{3}{2}}\frac{1}{m^{3}},
\end{equation}
to which corresponds to a characteristic mass of $M_{0}$:

\begin{equation}
M_{0}=\left( \frac{\hbar c}{G}\right) ^{\frac{3}{2}}\frac{1}{m^{2}},
\end{equation}
the model loses its validity. Everybody can recognize the fundamental
constant of the stellar systems, which has much to do with the stability
conditions of the stars (see in ref.\cite{chand}). Note that this constant
appears as consequence of the energy scaling assumed, so that, it can not be
obtained if a another scaling law had been adopted. A consequent analysis of
these massive systems should be performed taking into account the
relativisty effects.

Finally, let us remember the well-known white dwarfs model based on the
consideration of the Thomas-Fermi method to describe the state equation of
the degenerate nonrelativistic electronic gas, whose pressure supports the
hydrostatic equilibrium of the star. Using a simple dimensional analysis,
from this model can be easily derived the {\em same} scaling laws obtained
by us analysing the necessary conditions for the equivalence of the
statistical ensembles. This coincidence is not casual: for the
nonrelativistic particles these scaling laws only depend on the dimension of
the physical space.

Let us discuss now the results of the numerical calculations. In the FIG.5
it is shown the general dependency of the functions $h\left( \Phi
_{0}\right) $, $h_{1}\left( \Phi _{0}\right) $ and $h_{2}\left( \Phi
_{0}\right) $. Observe the oscillatory character of these functions with the
increasing of the parameter $\Phi _{0}$. This behavior is characteristic of
the isothermal distribution. It can also be seen the quasi-polytropic
asymptotic behavior for low values of $\Phi _{0}$, the power law $\Phi _{0}^{%
\frac{3}{4}}$.

\begin{figure}[tbp]
\caption{Behavior of the functions $h\left( \Phi _{0}\right) $, $h_{1}\left(
\Phi _{0}\right) $ and $h_{2}\left( \Phi _{0}\right) $. The oscillatory
behavior is characteristic of the isothermal distribution.}
\end{figure}

Throughout the Eq.(\ref{betequa}) it is obtained the dependence of the
canonical parameter $\beta _{o}$, as well as the energy $\epsilon $ from the
parameter $\Phi _{0}$. Similarly, it is studied this dependence for the
radio in which is contained the $80$ $\%$ of the total mass of the system, $%
R_{80\%}$, as well as for the central density $\rho _{0}$. All these
dependencies are shown in the FIG.6, 7, 8 and 9 respectively.

\begin{figure}[tbp]
\caption{Canonical parameter $\protect\beta $ vs $\Phi _{0}$. Observe the
persistence of the ascillatory behavior.}
\end{figure}

\begin{figure}[tbp]
\caption{Scaling invariant energy, $\protect\epsilon $ vs $\Phi _{0}$. It is
observe again the oscillatory behavior.}
\end{figure}

\begin{figure}[tbp]
\caption{Radio at the 80\% of the system total mass, $R_{80\%}$ vs $\Phi
_{0} $. }
\end{figure}

\begin{figure}[tbp]
\caption{Central density $\protect\rho _{0}$ vs $\Phi _{0}$. This function
grows monotonically with the increasing of $\Phi _{0}$. This result allows
us to understand that with the increasing of $\Phi _{0}$ the system develop
a core-halo structure (a high dense core and a dilute halo).}
\end{figure}

Let us to comment briefly these results. At the first place it is noted the
bounded character of the parameters $\beta _{o}$, $\epsilon $ and $R_{80\%}$
for all the values of the parameter $\Phi _{0}$: all of them are contained
in the following intervals:

\begin{equation}
\beta _{o}\in \left( 0,2.97\right) ;\text{ }\epsilon \in \left(
-0.26,0.055\right) ;\text{ }R_{80\%}\in \left( 1.02,1.33\right) .
\end{equation}
showing an oscillatory behavior for high values of $\Phi _{0}$ around the
values $\beta _{\infty }\simeq 2.12$, $\epsilon _{\infty }\simeq -0.18$, and 
$\left( R_{80\%}\right) _{\infty }\simeq 1.11$. However, the central density
grows monotonic with the increasing of $\Phi _{0}$, exhibiting an
exponential behavior for $\Phi _{0}>9.45$. It is very easy to understand
that our solution exhibits the same features of the Antonov problem solution 
\cite{antonov}. This is evident when analyzing the caloric curve: the
dependency $\beta _{o}$ vs $\epsilon $, the FIG.10: it is found again the
very well-known spiral. Additionally, it is shown\ in the FIG.11 the
dependence of the radio in which is contained the $80\%$ of the total mass
of the system with the energy.

\begin{figure}[tbp]
\caption{Caloric curve of the astrophysical system. It is obtained again the
classical spiral. In general, the results of the present model exhibit the
same features of the Antonov's problem.}
\end{figure}

\begin{figure}[tbp]
\caption{System size vs Energy. The system becomes smaller with the energy
decreasing.}
\end{figure}

All these configurations contained between the points {\bf B} and {\bf C}
can be described using the canonical description in the representation $%
\left( {\cal E},N\right) $. Here the system exhibits a positive heat
capacity. All the equilibrium configurations between the points {\bf C} and 
{\bf D} can not be accessed from the canonical description in the $\left( 
{\cal E},N\right) $ representation. In this part of the spiral the systems
exhibits a {\em negative heat capacity}. No equilibrium configurations exits
for values of $\beta _{o}$ greater than $\beta _{C}=2.97$. Ordinarily this
is referred as the {\em isothermal catastrophe}. On the other hand, no
equilibrium configurations exits for energies outside the interval $\left(
-0.26,0.055\right) $. For energies greater than $\epsilon _{B}=0.055$ the
system is extreme diluted and can not be confined. For energies below of $%
\epsilon _{D}=-0.26$, the systems collapse developing a core-halo structure.
It is obtained again the {\em gravothermal catastrophe}.

All those configurations contained between the points {\bf B} to {\bf D}
correspond to equilibrium situations, the other points of the spiral
correspond to unstable saddle points. This is evident when analyzing the
thermodynamic potentials of the ensembles, the FIG.12. and 13. For example,
in the FIG.12. it is represented the dependence of the Planck potential from
the canonical parameter $\beta _{o}$. Here it is evident that in the
canonical ensemble (using the $\left( {\cal E},N\right) $ representation)
all those configurations between the points {\bf B} to {\bf C }are stable: 
{\em in such points it is minimized this thermodynamic potential.}
Similarly, in the FIG.13. it is shown that in the microcanonical ensemble
all those configurations between the points {\bf B} to {\bf D }are stable: 
{\em in such points it is maximized the entropy for each accessible value of
the energy}.

\begin{figure}[tbp]
\caption{Planck potential vs canonical parameter $\protect\beta $. Here is
evident that in the canonical ensemble are stable all those configurations
between the points {\bf B} to {\bf C.}}
\end{figure}

\begin{figure}[tbp]
\caption{In the microcanonical ensemble are stable all those configurations
between the points {\bf B} to {\bf D.}}
\end{figure}

It has to be pointed out that the Gibbs' argument, the equilibrium of a
subsystem with a thermal bath, is not applicable to this situation, since it
is based on the independence or weak correlation of the subsystem with the
thermal bath. This is an invalid supposition for the nonextensive systems
due to the long-range correlations existing in them as a consequence of the
long-range interactions among the particles, which is the case that we are
studying here. It is needed to remember that the use of generalized
canonical description is supported by {\em its equivalency} with the
microcanonical description in the thermodynamic limit, which is the
description physically justified to this system. Moreover, no reasonable
thermal bath exits for the astrophysical systems. Thus, the isothermal
catastrophe is not a phenomenon with physical relevance since it can never
be obtained in nature: the consideration of a thermal bath in the
astrophysical system is out of context. A different significance possesses
the gravothermal catastrophe. The gravitational collapse is the main engine
of structuration in astrophysics and it concerns almost all scales of the
universe: the formation of planetesimals in the solar nebula, the formation
of stars, the fractal nature of the interstellar medium, the evolution of
globular clusters and galaxies and the formation of galactic clusters in
cosmology \cite{chava3}.

Many authors claim that those regions characterized by a negative heat
capacity can not be accessed by the canonical description, and the only way
to access to them is microcanonically \cite{gro2}. It should be remarked
that this assumption is not strictly correct. The reason is easy to
understand since the analysis of the equivalency between the canonical and
microcanonical ensemble does not need to be performed using a {\em global
representation} for the integrals of motions, the {\em local representations}
are admissible too. The microcanonical ensemble is local reparametrization
invariant: the description of an arbitrary small region of the integral of
motion space could be performed using any representation of the integrals of
motion: the physical observables in the microcanonical description do not
depend on the representation (see in ref. \cite{vel1}). Thats is the reason
why it is considered that the analysis of the equivalence of the
microcanonical ensemble with the canonical one does not have to be limited
to the use of only one representation for all the space (global analysis),
but this analysis could be made using a different representation in an
arbitrary small regions of the space (local analysis). In the previous
section it was used this argument to justified the validity of the Eq.(\ref
{betequa}) for all the accessible equilibrium configurations of the system,
showing in this way the importance of this geometrical characteristic of the
probabilistic distribution functions in the macroscopic description of
systems.

\section{Concluding remarks}

In the present paper it has been shown how the Boltzmann-Gibbs' Statistics
can be improved in order to extend its applicability to the study of the
macroscopic description of the {\em N}-body self-gravitating gas, although
this kind of systems are nonextensive. The Tsallis' Statistics is non
applicable to this kind of system since this theory demands potential
self-similarity scaling laws in the thermodynamic limit \cite{vel3}, which
is not the case that it is studying here: the astrophysical systems are
pseudoextensive \cite{vel2}, they exhibit an exponential scaling laws when $%
N\rightarrow \infty $.

The most important result obtained in the present analysis is the
specification of the thermodynamic limit for this kind of systems. Although
the {\em N}-independent picture of the traditional description of the
astrophysical systems remains inalterable, the bad consideration of the
scaling laws leads to a {\em trivial ensemble inequivalency}, which was
shown in the dynamical analysis performed in the ref. \cite{pet}.

The coincidences and connections of the results derived from this study with
others results obtained in the past using both, thermodynamic and non
thermodynamic methods, constitute important evidences of the validity of our
considerations.

Finally, it was shown the importance of the geometrical aspects of the
probabilistic distribution functions in the macroscopic study of systems.
Specifically, it was used these geometrical properties to extend the
validity of the generalized canonical description to those situations in
which the traditional description limited to the consideration of an special
representation for the space of the integrals of motion of the system fails.
These geometric aspects could be used to enhance the possibilities of the
Montecarlo method based on the canonical exponential weight during the study
of those hamiltonian systems presenting anomalies in their heat capacity.

\section{Appendix}

\subsection{Derivation of the descomposition formula.}

The $N$-body phase space ${\cal X}_{N}$ is an Cartesian external product of
the generalized coordinates and momentum spaces of each particle, $Q$ and $%
{\cal P}$, which can be represented in the following way:

\begin{equation}
{\cal X}_{N}=%
\mathrel{\mathop{\stackrel{N}{\prod }}\limits_{s=1}}%
\bigskip \left( Q_{s}\otimes {\cal P}_{s}\right) .
\end{equation}
Using the partition of the coordinates space\ $Q$ in not averlapped cells
\thinspace $\left\{ c_{k}\right\} $:

\begin{equation}
Q=%
\mathrel{\mathop{\bigcup }\limits_{k}}%
c_{k},
\end{equation}
the space ${\cal X}_{N}$ can be descomposed in not overlapped subspaces $%
{\cal X}_{N}^{\left( \sigma \right) }$:

\begin{equation}
{\cal X}_{N}=%
\mathrel{\mathop{\bigcup }\limits_{\sigma }}%
{\cal X}_{N}^{\left( \sigma \right) },  \label{desc 1}
\end{equation}
where 
\begin{equation}
{\cal X}_{N}^{\left( \sigma \right) }=%
\mathrel{\mathop{\stackrel{N}{\prod }}\limits_{s=1}}%
q_{s}\left( p_{\sigma }\left( s\right) \right) \otimes {\cal P}_{s}.
\end{equation}
Here, $p_{\sigma }\left( s\right) $ is a function which assigns each
particle of the system to a determined cell. The index $\sigma $ denotes all
the possible ways to perform this correspondence. Moreover, $q_{s}\left(
k\right) \equiv c_{k}$. Using the Eq.(\ref{desc 1}), the $N$-body phase
space integration can be expressed as follows:

\begin{equation}
\frac{1}{N!}%
\mathrel{\mathop{\int }\limits_{{\cal X}_{N}}}%
dX_{N}=%
\mathrel{\mathop{\sum }\limits_{\sigma }}%
\frac{1}{N!}%
\mathrel{\mathop{\int }\limits_{{\cal X}_{N}^{\left( \sigma \right) }}}%
dX_{N}.  \label{ec sum}
\end{equation}

Due to the identity of particles, all those terms corresponding to
configurations with identical occupation number of particles at the cells
are identical. Let $\left\{ n_{k}\right\} $ be the occupation numbers of
particles in the cells. For this this case, there are a total of:

\begin{equation}
C_{\left\{ n_{k}\right\} }^{N}=\frac{N!}{%
\mathrel{\mathop{\prod }\limits_{k}}%
n_{k}!}
\end{equation}
identical terms in the sum of the Eq.(\ref{ec sum}). Let ${\cal X}%
_{N}^{\left\{ n_{k}\right\} }$ be a subspace of ${\cal X}_{N}$ with
occupation numbers given by $\left\{ n_{k}\right\} $. Since ${\cal X}_{N}$
is a Cartesian external product of spaces, its volume element can be
factorized in the volume elements of the different spaces. This can be
convenientely done grouping all these particles beloging to the same cell.
In this case the $N$-body phase space integration of the subspace ${\cal X}%
_{N}^{\left\{ n_{k}\right\} }$ can be descomposed as follows:

\begin{equation}
\frac{1}{N!}C_{\left\{ n_{k}\right\} }^{N}%
\mathrel{\mathop{\int }\limits_{{\cal X}_{N}^{\left\{ n_{k}\right\} }}}%
dX_{N}=%
\mathrel{\mathop{\prod }\limits_{k}}%
\widehat{{\cal O}}\left[ {\cal X}_{n_{k}}^{\left( k\right) }\right] ,
\end{equation}
where:

\begin{equation}
\widehat{{\cal O}}\left[ {\cal X}_{n}^{\left( k\right) }\right] =\left\{ 
\begin{array}{c}
\frac{1}{n!}%
\mathrel{\mathop{\int }\limits_{{\cal X}_{n}^{\left( k\right) }}}%
dX_{n},\text{ if }n\neq 0, \\ 
1,\text{ if }n=0,
\end{array}
\right.
\end{equation}
and ${\cal X}_{n_{k}}^{\left( k\right) }$ represents the $n_{k}$-body phase
space whose physical space of the particles is reduced to the cell $c_{k}$.
Taking into account all the expossed above, it is easy to see that the $N$%
-body phase space integration can be finally expressed as:

\begin{equation}
\frac{1}{N!}%
\mathrel{\mathop{\int }\limits_{{\cal X}_{N}}}%
dX_{N}\equiv 
\mathrel{\mathop{\prod }\limits_{k}}%
\mathrel{\mathop{\stackrel{N}{\sum }}\limits_{n_{k}=0}}%
\widehat{{\cal O}}\left[ {\cal X}_{n_{k}}^{\left( k\right) }\right] \delta
^{\left( e\right) }\left( N-%
\mathrel{\mathop{\sum }\limits_{k}}%
n_{k}\right) ,  \label{desc fin}
\end{equation}
where the $\delta ^{\left( e\right) }\left( n\right) $ is related with the
Kronekel delta function as follows: 
\begin{equation}
\delta ^{\left( e\right) }\left( n\right) =\delta _{0n}.
\end{equation}
The presence of this function in the right hand of the Eq.(\ref{desc fin}),
assures that the number of particles remain fixed and equals $N$.

\bigskip

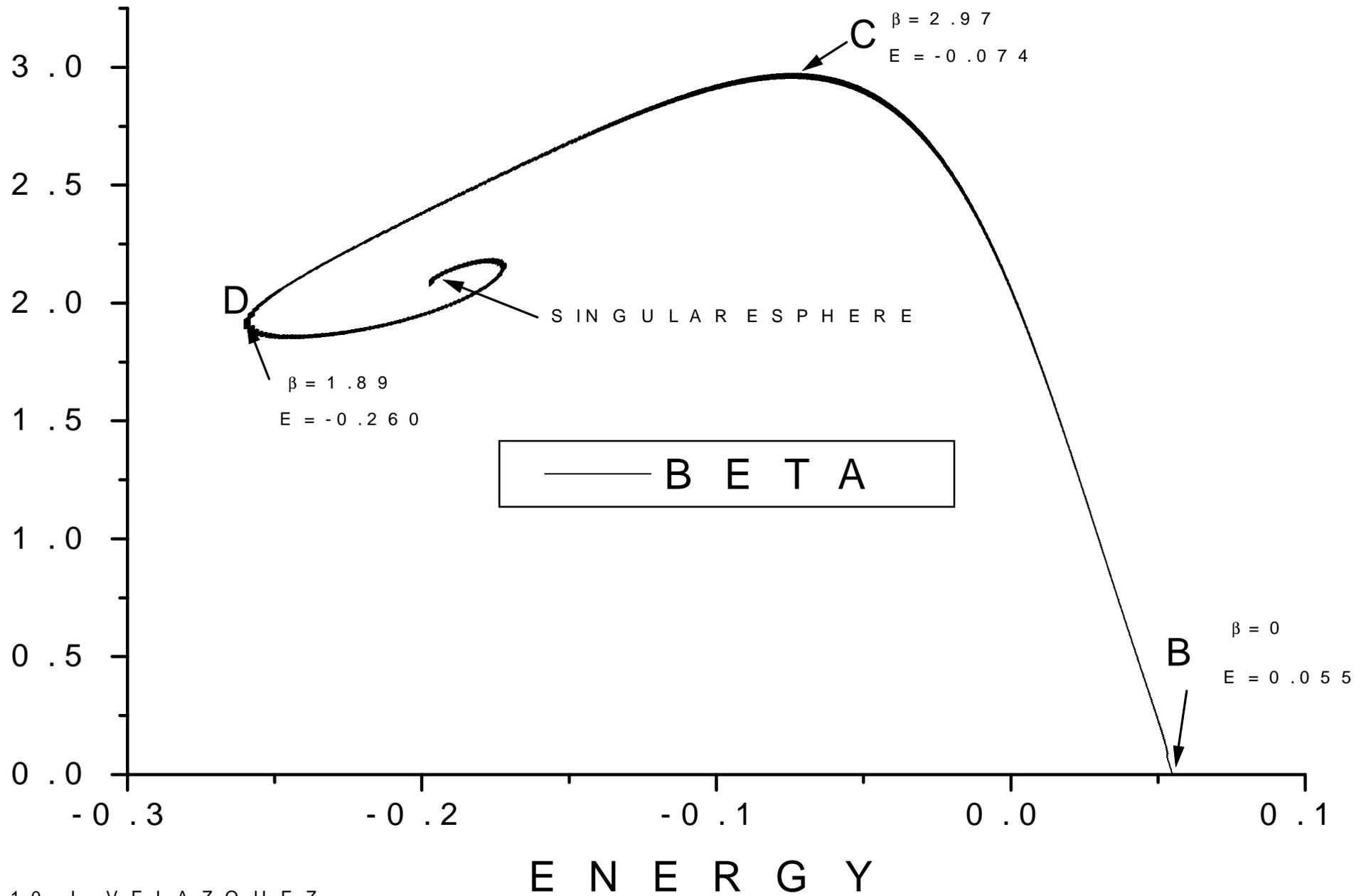
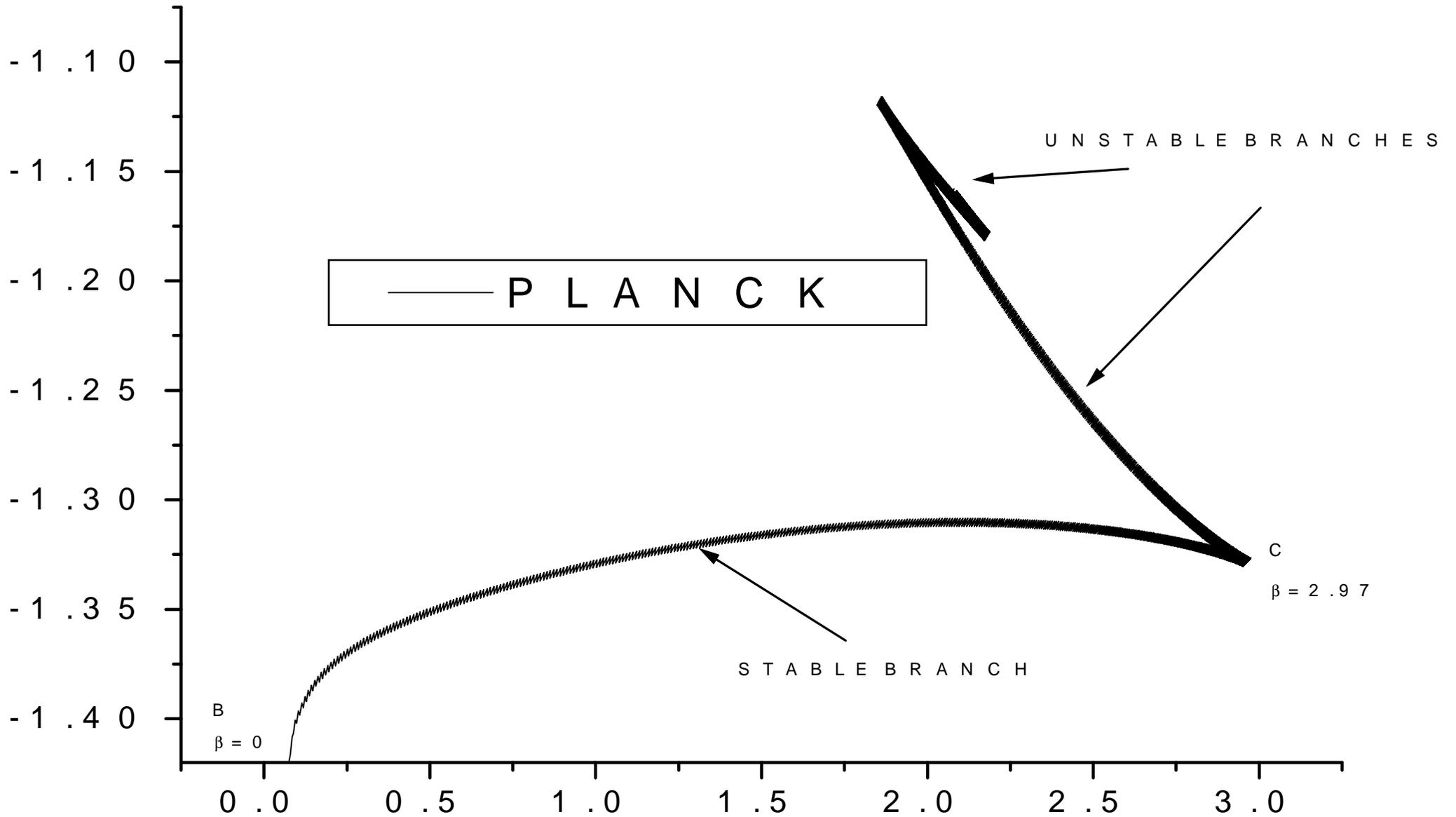
\end{document}